\documentclass[12pt,amssymb,amsmath]{article}
\usepackage{latexsym}
\usepackage{psfrag}

\usepackage{amssymb}
\usepackage{amsmath}
\usepackage{epsf}
\usepackage{epsfig}

    \newcommand{\Reals}{\it I\kern-.4emR}
    
    \newcommand{\Notin}{/\kern-.6em\hbox{$\in$}}
    \newcommand{\Notequiv}{/\kern-.6em\hbox{$\equiv$}}
    
    \newcommand{\Ceals}{\it I\kern-.65emC}

    \newcommand{\MM}{\it I\kern-.4emM}
    \newcommand{\NN}{\it I\kern-.4emN}
    \newcommand{\yy}{\it Y\kern-.8emY}
    \newcommand{\zz}{\makebox[.80em]{\it Z\kern-.46emZ}}
    \newcommand{\tzz}{\makebox[.80em]{\scriptsize\it Z\kern-.46emZ}}
    \newtheorem{theorem}{Theorem}[section]
    \newtheorem{lemma}[theorem]{Lemma}
    \newtheorem{fact}[theorem]{Fact}

    \newtheorem{claim}[theorem]{Claim}
    \newtheorem{observation}[theorem]{Observation}

\newcommand{\p}{^{\prime}}

\title{Characterizing 1-Dof Henneberg-I graphs with efficient
configuration spaces}
\author{Heping Gao$^{*}$,
Meera Sitharam
\footnote{
University of Florida; 
Work supported in part by NSF Grants
 EIA 02-18435, CCF 04-04116, and a Research Gift from SolidWorks}
}

\begin{document}

\maketitle
\thispagestyle{empty}

\begin{abstract}
We define and study exact, efficient representations of realization spaces
of a natural class of underconstrained 2D Euclidean Distance Constraint Systems(EDCS, Linkages, Frameworks)
based on 1-degree-of-freedom(dof) Henneberg-I graphs.
Each representation corresponds to a choice of parameters and
yields a different parametrized configuration space.
Our notion of efficiency is  based on the algebraic complexities of
sampling the configuration space and of obtaining a realization
from the sample (parametrized) configuration. Significantly, we give
purely combinatorial characterizations
that capture (i) the class of graphs
that have efficient configuration spaces
and (ii) the possible choices of representation parameters that yield efficient configuration
spaces for a given graph.
Our results automatically yield an efficient algorithm for
sampling realizations, without missing extreme or boundary realizations.
In addition, our results formally show that our definition of
efficient configuration space is
robust and that our characterizations are tight.  We choose the class of 1-dof Henneberg-I graphs in
order to take
the next step in a systematic and graded program
of combinatorial characterizations of efficient configuration spaces.
In particular, the results presented here are the first
characterizations that go beyond
graphs that have connected and convex configuration spaces.
\end{abstract}


\section{Introduction}
A {\em linkage} is a graph $G = (V,E)$ with fixed length
bars as the edges.   Denote by $\delta: E \rightarrow \mathbb{R}^1$  
the bar lengths.  The degrees of freedom ({\em dof}s) of a linkage
on the Euclidean plane refer to internal motions, after discounting Euclidean  
or rigid body motions that
rotate or translate the entire linkage, preserving all pairwise distances.
The problem of  describing the plane 
realizations   of {\em one degree-of-freedom linkages} or {\em mechanisms} 
has a long history.

A reasonable way to describe this space of realizations of a 1-dof linkage
$(G,\delta)$ is to take
a pair of vertices not connected by bars i.e, a 
{\em non-edge} $f$, and ask  for 
all the possible distance values $\delta^*$ that the non-edge $f$ can 
attain. 
This set of realizable distance values $\delta^*$ for the non-edge
$f$  is called the {\em configuration space of} the linkage $(G,\delta)$ 
on $f$, or parametrized by the distance $\delta^*(f)$.
This configuration space is a set of intervals on the real line.

For a well-known class of 1-dof linkages, we answer the following 
questions: How to describe the interval endpoints of 
such a configuration space? What is a reasonable and robust measure of complexity of this configuration space? Does the choice of non-edge $f$ influence
this complexity?  And
using such a complexity measure, which graphs $G$ have configuration
spaces of low complexity?

\subsection{Summary of Contributions}
Our class of 1-dof linkages is obtained from so-called
{\em Henneberg-I} graphs, a natural subclass of 
{\em Laman} or {\em minimally rigid} graphs. 
These graphs can be constructed one vertex at a time,
starting with a {\em base edge} $f$. At each step of the construction
a new vertex  is added with edges between it and 
exactly 2 previously constructed 
vertices.  See Figure~\ref{F:ste}. 
Delete the base edge $f$, denote the resulting 1-dof Henneberg-I 
graph as $G = (V,E)$, assign distances $\delta$ to the edges to obtain a 
1-dof linkage $(G,\delta)$.

Denote the configuration space of this linkage $(G,\delta)$ on $f$
as $\Phi_f(G,\delta)$. As mentioned earlier, this is a set of intervals.
Given an configuration $\delta^*$ in this set, 
a corresponding cartesian realization  - 
which assigns the distance value $\delta^*$ to $f$ - can be computed using a 
ruler and compass:  simply follow the partial 
order of the Henneberg construction, and realize each 
vertex as a point in $\mathbb{R}^2$,
by solving one quadratic equation in one variable
at each step.

Algebraically, this is the solution of a triangularized system of
quadratics the complexity of which is generally refered to as
{\em Quadratic or Radical Solvability.}

More specifically, we answer the following questions.

\begin{itemize}
\item [(1)]
 What do the endpoints of the intervals in the set $\Phi_{f}(G,\delta)$ above
correspond to? We show in Theorem \ref{thm:config-space} 
that they have a combinatorial meaning, in
fact, they can be computed by realizing other linkages, called {\em extreme}
linkages obtained from 
the graph $G$ and the non-edge $f$.

\item[(2)]
For which $G$ and $f$ is the complexity of obtaining 
endpoints of the above intervals roughly the same as the 
ruler and compass realization 
complexity described above? 
More precisely, we use (1) and ask when are all the extreme
linkages Quadratically solvable?

\begin{figure}[h]
\psfrag{1}{$v_1$}
\psfrag{2}{$v_2$}
\psfrag{3}{$v_3$}
\psfrag{4}{$v_4$}
\psfrag{5}{$v_5$}
\psfrag{6}{$v_6$}
\psfrag{7}{$v_7$}
\psfrag{d13}{$7$}
\psfrag{d23}{$7$}
\psfrag{d14}{$6$}
\psfrag{d24}{$8$}
\psfrag{d35}{$4.5$}
\psfrag{d45}{$0.5$}
\begin{center}
\includegraphics[width=12cm]{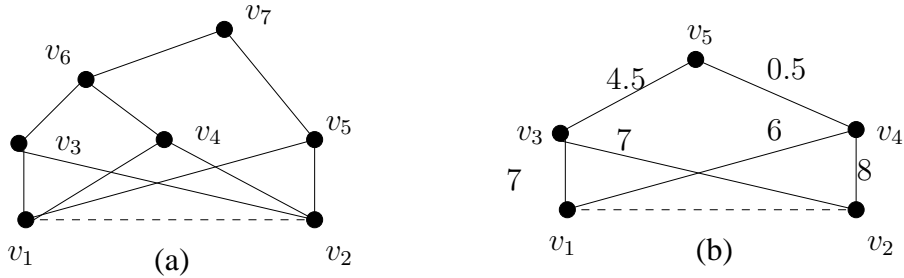}
\end{center}
\caption{
Figure (left) is a 1-Dof Henneberg-I graph whose configuration space on the
base non-edge has interval endpoints that are not always quadratically solvable.
Figure (right) on the other hand has quadratically solvable end points. 
For the edge distances show, the intervals are 
$[\frac{1}{8}\sqrt{6214-90\sqrt{17}\sqrt{209}},\frac{1}{8}\sqrt{6214+6\sqrt{17}\sqrt{209}}]$
and 
$[\frac{2}{5}\sqrt{565-360\sqrt{2}},\frac{2}{5}\sqrt{565+360\sqrt{2}}]$.
}
\label{fig:initial}
\end{figure}

Figure \ref{fig:initial} shows two examples of 
graphs $G$ and non-edges $f$: 
the interval endpoints 
are quadratically solvable for one of them, but not 
for the other.

In fact, we ask for which $G$ and $f$,
the extreme graphs have a graph property called
{\em Tree-} or {\em Triangle decomposability}, which has been shown in 
\cite{bib:Owen02}
to be generically equivalent to Quadratic Solvability 
for planar graphs and the equivalence is strongly conjectured for all graphs.
We say that such configuration spaces $\Phi_f(G,\delta)$
have {\em low sampling complexity}.
We give in Theorem \ref{the:TriangleFreeCase} 
a forbidden minor characterization of the property of low sampling complexity
and in Observation \ref{obs:backward}
give a faster algorithm for
finding the interval endpoints in $\Phi_f(G,\delta)$ than by realizing all the
extreme graphs as per (1). We also 
show in Observations \ref{obs:TriangleFreeCounter}, \ref{obs:GeneralTriCounter} and \ref{obs:onepathcounter} 
the tightness of this forbidden minor 
characterization by dropping various conditions and showing that no
forbidden minor characterization will apply. Furthermore, in 
Theorem \ref{the:chain}, we give an 
algorithmic characterization for a larger class of graphs.

\item[(3)]
 Does the choice of the base non-edge $f$  matter ?
A Henneberg-I graph could be constructible from different possible
base edges and a 1-dof Henneberg-I graph  could be obtained 
by deleting any one of them.
Could these configuration spaces have different sampling complexities?

In Theorem \ref{the:quantifierExchange} 
show that this cannot happen, thereby showing that our measure
of sampling complexity for configuration spaces of  1-dof Henneberg-I
linkages is robust.
\end{itemize}

\subsection{Model of Computation}
Our complexity measures are based on a model of computation
that uses exact representation of numbers in any 
quadratic extension field of the rational numbers.
In other words, we assume that all arithmetic operations, comparisons
and extraction of square roots are constant time, exact operations.
This model of computation is not as strong as the real 
RAM model that is normally used in computational geometry, 
that permits exact representation of arbitrary algebraic 
numbers \cite{bib:Loos}. 
Issues in exact geometric computation  such as efficient and robust 
implementation of such a representation,  for example using interval 
arithmetic,  are beyond the scope of this manuscript.

\subsection{Organization}

In Section \ref{sec:motivation} 
we motivate and give a brief background for the overall 
program of investigation including various measures of efficiency
of configuration spaces.
The contributions of this manuscript are aligned with this program. 
Their novelty and technical significance is outlined in 
Section \ref{sec:novelty} together with related work. 
The theorems and proofs are presented in Section~\ref{sec:results}.
We conclude with suggestions for future work in Section \ref{sec:conclusion}.

\section{Overall Program and  Motivation} 
\label{sec:motivation}

We begin by clarifying and unifying terminology that arises in different
communities that are interested in the same problems concerning
configuration spaces of linkages.  In geometric constraint solving
terminology, a  linkage is also called 
a {\em Euclidean Distance Constraint System (EDCS)} $(G,\delta)$, i.e., 
is a graph $G = (V,E)$ together with an assignment of distances
$\delta(e)$, or distance intervals $[\delta^l(e),\delta^r(e)]$ to
the edges $e \in E$.  A $d$-dimensional {\em realization} is the
assignment $p$ of points in $\mathbb{R}^d$ to the vertices in $V$
such that the distance equality (resp. inequality) constraints
are satisfied: $\delta(u,v) = \|p(u) - p(v)\|$ (respectively
$\delta^l(u,v) \le \|p(u)-p(v)\| \le \delta^r(u,v)$). Note: an
EDCS with distance equality constraints, $(G,\delta)$, was
originally refered to as a {\em framework} in combinatorial
rigidity terminology; more recently a framework $(G,p)$ includes
a specific realization $p$, and the distance assignment $\delta$
is read off from $p$. 

{\bf Note:} 
We will use standard and well-known 
geometric constraint solving 
(and the corresponding combinatorial rigidity) 
terminology for which we refer the
reader to, for example \cite{bib:survey} \cite{bib:FudHo97} and
\cite{bib:Graver}.
In 2D, a  graph $G=(V,E)$ is
{\em wellconstrained or minimally rigid} if it satisfies the {\em Laman} conditions
\cite{bib:Laman70}; i.e.,  $|E|=2|V|-3$ and
$|E_{s}| \leq 2|V_{s}|-3$ for all subgraphs $G_{s}=(V_{s},E_{s})$ of $G$;
$G$ is {\em underconstrained} or {\em independent and not rigid} if we have $|E| < 2|V|-3$
and $|E_{s}| \leq 2|V_{s}|-3$ for all subgraphs $G_{s}$. A graph $G$
is {\em overconstrained or dependent} if there is a subgraph $G_{s}=(V_{s},E_{s})$
with $|E_{s}| > 2|V_{s}|-3$.
$G$ is {\em welloverconstrained or rigid} if there exists a subset of its edges $E'$
such that the graph $G' = (V,E')$
is wellconstrained or minimally rigid.
A graph is {\em flexible} if it is not rigid.

One seeks {\sl efficient representations of the realization space}
of an EDCS. We define a {\em representation} to be (i) a choice
of parameter set, specifically a choice of a set $F$ of non-edges
of $G$, and (ii) a set $\Phi_F^d(G,\delta)$ of possible distance
values $\delta^*(f)$ that the non-edges in $f\in F\subseteq
\overline{E}$ can take while ensuring existence of at least one
$d$-dimensional realization for the augmented EDCS: $(G\cup F,
\delta(E),\delta^*(F))$. Here $G \cup F$ refers to a graph $H :=
(V, E\cup F)$. In other words, the 
representations employ {\em Cayley} parameters: 
distances or sometimes {\sl squared} distances corresponding to the 
non-edges in $F$ \cite{bib:cayley}.  The set
$\Phi_F^d(G,\delta)$ is the {\sl projection} of the Cayley-Menger
semi-algebraic set associated with $(G,\delta)$ on the Cayley
parameters in $F$. As mentioned earlier, 
we refer to the representation $\Phi_F^d(G,\delta)$ as the {\em
configuration space} of the EDCS $(G,\delta)$ {\em on the
parameter set $F$} of non-edges of $G$.

Describing and sampling the realization space of an EDCS is a
difficult problem that arises in many classical areas of
mathematics and theoretical computer science and has a wide
variety of applications in computer aided design for mechanical
engineering, robotics and molecular modeling. Especially 
for {\it underconstrained} ({\it independent} and not {\it rigid}) EDCS whose realizations have 
one or more internal degrees of freedom of motion, progress
on this problem has been very limited.

Existing methods for sampling EDCS realization spaces
often use Cartesian representations, 
factoring out the Euclidean group by 
arbitrarily ``pinning" or ``grounding"
some of the points' coordinate values. 
Even when the methods use internal representation
parameters such as Cayley parameters (non-edges) 
or angles between unconstrained objects, the choice of these 
parameters is usually adhoc.
    While Euclidean motions are automatically factored out in the 
    resulting parametrized configuration space, 
for most such parameter choices, the  parametrized
configuration space is still a topologically complex
semi-algebraic set, often of reduced measure in high dimensions. 
The method of sampling
is usually:  ``take a uniform grid sampling and 
throw away sample configurations that do not
satisfy constraints." 
Since even configuration spaces of full measure 
(representation with lowest possible 
number of parameters or dimensions) often have complex boundaries,
this type of sampling method 
is likely to  miss extreme and boundary
configurations and is moreover computationally inefficient. 
To deal with this,  numerical, iterative methods are generally
used in case that
the constraints are equalities, and in the case of inequalities, 
probabilistic ``roadmaps" and
other general collision avoidance methods are used. They
are approximate methods that do not leverage exact descriptions of the 
configuration space. 


Two related problems additionally occur in NMR molecular
structure determination and wireless sensor network localization:
completing a partially specified Euclidean Distance Matrix in a
given dimension; and finding a Euclidean Distance Matrix in a
given dimension that closely approximates a given Metric Matrix
(representing pairwise distances in a metric space)  
\cite{bib:alfakih99,bib:yinyuye06,bib:havel88}.
The latter
problem also arises
in the
study of
algorithms for low distortion embedding of metric spaces into
Euclidean spaces of fixed dimension \cite{bib:indyk06}. Both of
these
problems can in fact directly be viewed as searching over a
configuration space of an EDCS.


\subsection{Exact, efficient configuration spaces}
\label{sec:efficient}

Motivated by these applications, our emphasis is on {\sl exact,
efficient} configuration spaces for underconstrained EDCS. First,
an exact algebraic description, 
given by polynomial inequalities - whose coefficients are  obtained 
after performing algebraic computations on the given EDCS - guarantees
that boundary and extreme configurations are not missed during
sampling, which is important for many applications.

{\sl Efficiency} refers to several factors.  We list four
efficiency factors. The
first factor is the {\em sampling complexity}: given the EDCS
$(G,\delta)$, (i) the complexity of computing (ia) the set of Cayley
parameters or non-edges $F$ and (ib) the description of the
configuration space $\Phi_F^d(G,\delta)$ as a 
semi-algebraic set, which includes the algebraic complexity of 
the {\sl coefficients} in the polynomial inequalities that describe the 
semi-algebraic set,
 and (ii) the descriptive 
algebraic complexity, i.e., number, terms, degree 
etc of the polynomial inequalities that describe the 
semi-algebraic set.
These together determine the complexity of
sampling or walking through configurations in
$\Phi_F^d(G,\delta)$.
            
Concerning (i) it is important to note that most choices of
Cayley parameters (non-edges) to represent the realization space
of $(G,\delta)$ give inefficient descriptions of the resulting
parametrized configuration space. Hence a strong
emphasis needs to be placed on a {\sl systematic, combinatorial choice} of the
Cayley parameters that {\sl guarantee} a configuration space with
{\sl all} the efficiency requirements listed here. Further, we
are interested in combinatorially characterizing {\sl for which
graphs $G$ such a choice even exists}.

The second efficiency factor is the {\em realization complexity}.
Note that the price we pay for insisting 
on exact and efficient configuration spaces
is that the map from the traditional Cartesian realization space
to the
parametrized configuration space is many-one. 
I.e, each parametrized configuration could correspond to many
(but at least one) Cartesian realizations.

However, we circumvent this difficulty by defining and studying
{\em realization complexity} as one of the
{\sl requirements} on efficient configuration spaces
i.e., we take into account that the realization step
typically follows the sampling step, and
ensure that one or all of the corresponding Cartesian
realizations can be obtained efficiently from a parametrized
sample configuration.

A third efficiency factor is {\em generic completeness}, i.e, we
would like each configuration in our parametrized configuration
space to generically correspond to at most finitely many
Cartesian realizations and moreover, we would like the
configuration space to be of {\em full-measure}, i.e, use exactly
as many parameters or dimensions as the internal degrees of
freedom of $G$.  Combinatorially this means that the graph $G\cup
F$ is well-constrained or minimally rigid in combinatorial
rigidity terminology.

A fourth important efficiency factor is {\em topological
complexity} for example, connectedness or number of connected components and 
{\em geometric complexity} for example convexity; however, for this 
manuscript, these factors are subsumed in the {\em sampling complexity} since
configuration space of this manuscript is a 1-parameter space.

In \cite{bib:Gao08} and \cite{bib:GaoSitharam08a} 
a series of exact combinatorial characterizations are given for 
connected, convex and complete configuration spaces of low sampling
and realization complexity for general 2D and 3D
EDCSs (including distance inequalities), 
and a somewhat weaker characterization is given for arbitrary dimensional
EDCSs.

\subsection{Combinatorial Characterization}
Combinatorial characterizations of generic properties of EDCS
are the cornerstone of combinatorial rigidity theory. In practice
they crucial for tractable and efficient geometric constraint solving, since
they are used to analyze and decompose the underlying algebraic system.
So far such characterizations have been used primarily for broad
classifications
into well- over- under- constrained, detecting dependent constraints
in overconstrained systems and finding completions for
underconstrained systems. Such combinatorial characterizations have been
missing in the finer classification of underconstrained
systems according to the efficiency or complexity of their
configuration space. This
however is a crucial step in efficiently decomposing and analyzing
underconstrained systems.
Our emphasis in this respect is the surprising fact that there is a
clean combinatorial characterization {\sl at all} of the algebraic
complexity of configuration
spaces.

The PhD thesis \cite{bib:Gao08} formulates the concept of 
efficient configuration space description for underconstrained EDCS, by
emphasizing the systematic choice of parameters that yield
efficient representations of the realization space, setting the
stage for a mostly combinatorial, and complexity-graded
program of investigation.  An initial sketch of this program was
presented in \cite{bib:GaoSi05};  a comprehensive list of
theoretical results and applications to date can be found in the
PhD thesis \cite{bib:Gao08}.
In this manuscript, we take the first step in one 
of two natural directions
to move beyond \cite{bib:GaoSitharam08a} which 
characterizes graphs whose EDCS always admit convex and/or 
connected 2D configuration spaces. 
One possible extension direction is to ask which graphs always admit
2D configuration spaces with at most 2 connected components.
Results in this direction can be found in \cite{bib:Gao08}.
A second possible direction, is to take the 
simplest natural class of graphs with 1-dof 
(generic mechanisms with 1-degree-of-freedom) that do {\sl not} have connected
configuration spaces, and combinatorially classify them based on 
their sampling complexity. This is the direction we take here.

\section{Novelty and Related Work}
\label{sec:novelty}


Our results give a practically meaningful, and mathematically robust 
definition of efficient configuration spaces for
a natural class of 1-dof linkages or EDCS, 
based on algebraic complexity of 
sampling and realization. Significantly, we give  
purely combinatorial, tight characterizations
that capture (i) the class of EDCS that have such configuration spaces 
and (ii) the possible choices of parameters that yield such configuration spaces.

To the best of our knowledge, the only known result in this area 
that has a similar flavor of 
combinatorially capturing algebraic complexity is the result
of \cite{bib:Owen02} that relates quadratic solvability and Tree- or Triangle-
decomposability for planar graphs.

Concerning the use of Cayley parameters or non-edges
for parametrizing the configuration space:  the papers 
\cite{bib:JoanArinyo03}, 
\cite{bib:survey} and \cite{bib:ZhangGao06}
study how to obtain ``completions'' of underconstrained graphs $G$, i.e, a set
of non-edges $F$ whose addition makes $G$ well-constrained or minimally rigid.
All are motivated by the need to efficiently 
obtain realizations of underconstrained
EDCS.  In particular \cite{bib:JoanArinyo03} also guarantees that the completion
ensures Tree- or Triangle- decomposability, thereby ensuring low realization
complexity.

However, they do not even attempt to address
the question of how to find realizable distance {\sl values} for
the completion edges.
Nor do they concern  themselves with
the geometric, topological or algebraic complexity of
{\sl the set of distance values
that these completion non-edges can take}, nor the complexity of obtaining
a description of this configuration space,
given the EDCS $(G,\delta)$ and the non-edges
$F$, nor a combinatorial characterization of graphs for which this
sampling complexity is low.
The latter factors however  are crucial for tractably analyzing and
decomposing underconstrained systems and for
sampling their configuration spaces
{\sl in order to} obtain the corresponding realizations.
The problem has generally been considered too messy, and
there has been no systematic, formal program to study this problem.

On the other hand, \cite{bib:hilderick06} gives a collection of useful 
observations and heuristics for 
computing the interval endpoints in the configuration space descriptions of certain
graphs that arise in real 
CAD applications.

\section{Results}
\label{sec:results}
\subsection{Definition and basic properties of Simple 1-dof Henneberg-I graphs}
\label{subsec:1dofHen1}

\begin{figure}[h]
\psfrag{1}{$v_1$}
\psfrag{2}{$v_2$}
\psfrag{3}{$v_3$}
\psfrag{4}{$v_4$}
\psfrag{5}{$v_5$}
\psfrag{6}{$v_6$}
\psfrag{7}{$v_7$}
\begin{center}
\includegraphics[width=12cm]{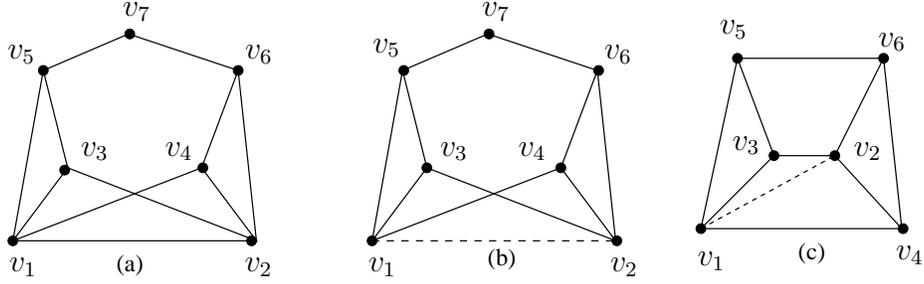}
\end{center}
\caption{
(a) Henneberg-I graph: ($v_1,v_{2})$ is the base edge;
(b) Simple 1-dof Henneberg-I graph: $(v_1,v_{2})$ is the base 
non-edge; (c) The extreme graph of (b) that corresponds to 
$v_{7}\triangleleft (v_{5},v_{6})$; it is also a
$K_{3,3}$ graph. For both (a) and (b), the 
Henneberg-I constructions contain 
$(v_{3}\triangleleft (v_1,v_{2}),v_{4}\triangleleft
(v_1,v_{2}),v_{5}\triangleleft
(v_1,v_{3}),v_{6}\triangleleft (v_{2},v_{4}),v_{7}\triangleleft
(v_{5},v_{6}))$.
}
\label{F:ste}
\end{figure}

As mentioned in the Introduction, 
{\em Henneberg-I} graphs can be constructed one vertex at a time,
starting with a {\em base edge}. At each step $k$ of the construction
a new vertex $v_k$ is added with edges to exactly 2 previously constructed vertices $u,w$,
called the {\em base pair of vertices at step $k$}. 
We denote this by $v_k \triangleleft u,w$. 
In fact, a Henneberg-I construction $c$ is 
actually  a  {\sl partial order} that is completely specified
by the base edge $f$, although we loosely use the phrase
{\em construction sequence} to refer to this partial order. 
See Figure~\ref{F:ste}.  We consider this class because it is the smallest 
natural class that contains 
2-trees (sometimes called graphs of tree-width 2) which figure prominently in 
the combinatorial characterizations of convex and connected configuration spaces
for 2D EDCS in \cite{bib:Gao08, bib:GaoSitharam08a}, as mentioned in
Section \ref{sec:motivation}.  
In other words,  Henneberg-I graphs are   
the simplest generalization of 2-trees which do not have 
convex or connected configuration spaces.
Henneberg-I graphs are a natural subclass of Laman or minimally rigid
graphs, and also of another common 
class of graphs called {\em Tree- or Triangle- decomposable}
graphs\cite{bib:FudHo97}, that are conjectured to be exactly equivalent
to quadratically solvable graphs, a conjecture that has been
proven for planar \cite{bib:Owen02}.

\begin{figure}[h]
\psfrag{A}{$v_1$}
\psfrag{B}{$v_2$}
\psfrag{C}{$v_3$}
\psfrag{1}{$G_1$}
\psfrag{2}{$G_2$}
\psfrag{3}{$G_3$}
\begin{center}
\includegraphics[width=4.5cm]{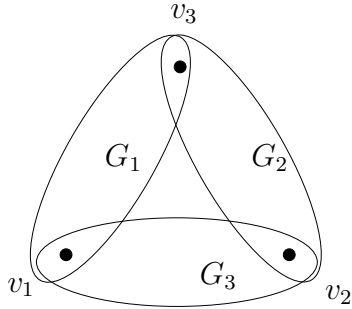}
\end{center}
\caption{
Tree-Decomposable Graph: a graph $G$ is Tree-Decomposable if it can be
divided into three Tree-Decomposable subgraphs $G_{1}$, $G_{2}$ and $G_{3}$
such that $G = G_{1} \cup G_{2} \cup G_{3}$, $G_{1} \cap G_{2} = (\{v_3\},
\emptyset)$, $G_{2} \cap G_{3} = (\{v_2\}, \emptyset)$ and $G_{1} \cap G_{3} =
(\{v_1\}, \emptyset)$ where $v_1$, $v_2$ and $v_3$ are three different vertices; as
base cases, a pure edge and a triangle are defined to be Tree-Decomposable.
}
\label{F:triangleDecomposition}
\end{figure}

A graph $G$ is {\em Triangle-Decomposable} or {\em Tree-Decomposable}, if:
\begin{itemize}
\item it is a pure edge or a triangle;
or
\item 
it can be divided into three Triangle-Decomposable subgraphs $G_{1}$,
$G_{2}$ and $G_{3}$ such that $G = G_{1} \cup G_{2} \cup G_{3}$, $G_{1} \cap
G_{2} = (\{v_3\}, \emptyset)$, $G_{2} \cap G_{3} = (\{v_2\}, \emptyset)$ and $G_{1}
\cap G_{3} = (\{v_1\}, \emptyset)$ where $v_1$, $v_2$ and $v_3$ are three different 
vertices (refer to Figure~\ref{F:triangleDecomposition})\cite{bib:FudHo97}.
\end{itemize}
We also say $G_1$, $G_2$ and $G_3$ are {\em clusters} and $v_1$, $v_2$ and
$v_3$ are {\em shared vertices}.

A generalization of the results presented here
from Henneberg-1 graphs to the larger class of Tree- or 
Triangle-decomposable graphs
appears in \cite{bib:Gao08} and \cite{bib:GaoSitharam08b}.

\begin{figure}[h]
\psfrag{1}{$v_1$}
\psfrag{2}{$v_2$}
\psfrag{3}{$v_3$}
\psfrag{4}{$v_4$}
\psfrag{5}{$v_5$}
\psfrag{6}{$v_6$}
\psfrag{7}{$v_7$}
\begin{center}
\includegraphics[width=12cm]{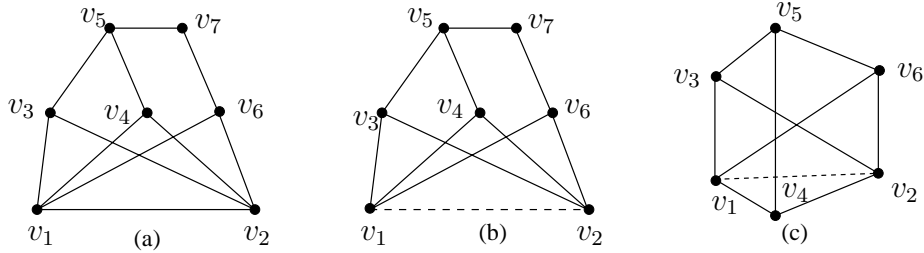}
\end{center}
\caption{
(a) Henneberg-I graph: $(v_1,v_{2})$ is the base edge;
(b) Simple 1-dof Henneberg-I graph: $(v_1,v_{2})$ is the base
non-edge; (c) The extreme graph of (b) that corresponds to
$v_{7}\triangleleft (v_{5},v_{6})$; it is also  a
$C_{3}\times C_{2}$ graph. For both (a) and (b), the 
Henneberg-I constructions contain
$(v_{3}\triangleleft (v_1,v_{2}),v_{4}\triangleleft
(v_1,v_{2}),v_{5}\triangleleft (v_{3},v_{4}),v_{6}\triangleleft
(v_1,v_{2}),v_{7}\triangleleft
(v_{5},v_{6}))$.
}
\label{F:hex}
\end{figure}

A {\em Simple 1-dof Henneberg-I} graph $G$ is obtained by removing a
base edge $f$ from a Henneberg-I graph (note that there can be more
than 1 possible base edge for a given Henneberg-I graph, refer to
Figure~\ref{F:OnePathCounter}).  Such an edge $f$ is called a {\em
base non-edge} of $G$. The EDCSs $(G,\delta)$ based on such graphs
generically have one internal degree of freedom and hence a {\sl
complete, 1-parameter} configuration space.

The notion of an {\em extreme graph} of a Simple 1-dof Henneberg-I
graph $G$ with base non-edge $f$ will be used prominently
in our results. The $k^{th}$ {\em extreme graph} $X_k$ based on $G$ and $f$
is obtained from $G$ by adding a new edge $(u,w)$ between the base pair of vertices $u$ and $w$
of the $k^{th}$ Henneberg construction step $v_k\triangleleft u,w$, provided
$u,w$ do not belong to any well-constrained subgraph of $G$  (otherwise, the
$k^{th}$ extreme graph  is overconstrained and irrelevant - 
depending on the context it could be left undefined).
For the linkage or EDCS $(G,\delta)$ and the non-edge $f$, 
the $k^{th}$ {\em extreme linkage or EDCS}  $X_{k,j}, j = 1,2$ is 
$(X_k, \delta^j)$, where the $j = 1,2$ represents two possible
extensions of $\delta$ to the new edge $(u,w)$:
$\delta^1(u,w) := \delta(u,v_k) +\delta(v_k,w)$, and   
$\delta^2(u,w) := |\delta(u,v_k) -\delta(v_k,w)|$.

Next we prove a series of facts giving basic
properties of 1-dof Henneberg-I graphs that 
will be used in our main results and are additionally of 
independent interest since these graphs are commonly occuring.

\begin{fact} 
\label{fact:wellcondition}
No subgraph of a Simple 1-dof Henneberg-I graph is
overconstrained (i.e., it is independent).
A subgraph $G'$ of a Simple
1-dof Henneberg-I graph is wellconstrained (minimally rigid) 
if and only if $G'$ is a
Henneberg-I graph. 
\end{fact}

\noindent
\begin{proof}
First we prove that no subgraph of a Simple 1-dof Henneberg-I graph is
overconstrained by showing that such a graph $G$ 
satisfies the Laman sparsity or independence condition \cite{bib:Laman70}:
i.e, the number of edges of any subgraph is at most twice the number of 
vertices minus 3. 
First, we consider the list of 
vertices of $G$ obtained from a 
Henneberg-I construction sequence $s$ for $G$ with base non-edge $f$. 
That is, $s$ is a 
Henneberg-I construction sequence 
for $G \cup f$, starting from $f$.
We start the list with the two vertices of $f$, 
(any relative ordering of these two vertices is fine).
Then, we add all the other
vertices to the list one by one strictly following the 
construction sequence $s$. One property of Henneberg-I sequences is that 
any vertex not in the first two slots of the list is adjacent to exactly 
two vertices which are before it in the list. 
For any subgraph $G'$, we can get a new list by
extracting the sublist corresponding to the vertices of $G'$ from this
list. 
In this sublist, any vertex is adjacent to at most 
two vertices which are before it.
Therefore, if the number of the vertices of $G'$ is 
$n$, the number of the edges of $G'$ will not exceed
$1 +2(n-2) = 2n -3$, thus ensuring the Laman sparsity or independence condition. 

Then we prove that a subgraph $G'$ of a Simple
1-dof Henneberg-I graph is wellconstrained (minimally rigid) 
if and only if $G'$ is a
Henneberg-I graph. 
One direction is clear since any Henneberg-I graph is wellconstrained.

For the other direction, by Laman's theorem~\cite{bib:Laman70}, 
the number of edge of $G\p$ has to be $2n -3$ if $G\p$ is 
wellconstrained. We have just proved that the number of edge
in $G\p$ does not exceed $2n -3$. For the equality to be true, 
there must be one edge between the first two vertices in the
sublist and
any vertex in the third or higher slot in the sublist must be adjacent to exactly two
vertices before it in the sublist. 
By the definition of Henneberg-I graph, this implies $G'$ has to be a Henneberg-I graph.
\end{proof}

\begin{fact} 
\label{fact:noWellOn12}
Given a Simple 1-dof Henneberg-I graph $G$ with base non-edge $f = (v_1,v_{2})$,
no wellconstrained subgraph $G'$ of $G$ can 
contain both $v_1$ and $v_2$.
\end{fact}

\noindent
\begin{proof}
The contrapositive follows from  
Fact \ref{fact:wellcondition} and its proof. 
That lemma states that $G'$ must 
be a Henneberg-I graph if it is wellconstrained and its proof points out that 
if $G'$ contains $v_1$ and $v_2$, there must be an edge 
the first two vertices in the sublist which are
$v_1$ and $v_2$ here. This contradicts $(v_1, v_2)$ being
the base non-edge of $G$.
\end{proof}

\begin{fact} 
\label{fact:extremeWell}
Take a Simple 1-dof Henneberg-I graph $G=(V,E)$ with a base non-edge
$(v_1,v_2)$
and corresponding Henneberg-I construction sequence
$(v_3 \triangleleft (u_3,w_3),\cdots,v_{n} \triangleleft (u_{n},w_{n})$
where $n=|V|$. Then
\begin{enumerate}
\item
 for any $m$, the
{\em extreme graph}
corresponding to $v_m \triangleleft (u_m,w_m)$, 
i.e., the graph obtained by adding the edge $(u_m,w_m)$ 
is wellconstrained if and only if
there is no wellconstrained subgraph in $G$ that contains both $u_m$ and
$w_m$.
\item
If there exists a subgraph $G'$ containing $u_m$ and $w_m$ that is 
wellconstrained, then we can say the following. Taking $G_{m-1}$ to be
the graph constructed before $v_m$ and 
let $G_m = G_{m-1} \cup v_m$
Now for any distance assignment $\delta$ we have 
$\Phi_f^2(G_m,\delta)= \Phi_f^2(G_{m-1},\delta)$ or
$\Phi_f^2(G_m,\delta)= \emptyset$.
\end{enumerate}
\end{fact}

\noindent
\begin{proof}
We first prove (1).
If there is a wellconstrained subgraph $G'$ containing both $u_m$ and
$w_m$, then $G' \cup (u_m,w_m)$ will be overconstrained. This 
proves one direction. For the other direction, if there is no
wellconstrained subgraph $G'$ containing both $u_m$ and
$w_m$, $G \cup (u_m,w_m)$ will not have any overconstrained subgraphs;
and since $G$ is 1-dof, $G \cup (u_m,w_m)$ would be 
wellconstrained (both by Laman's theorem \cite{bib:Laman70}). 
This proves the other direction. 

For (2), by Fact~\ref{fact:wellcondition}, $G'$ is a Henneberg-I graph
with a base edge, say $(v_i,v_j)$. If we remove all 
the vertices of $G\p$ other than $v_i$ and $v_j$.
we can get a subgraph $G^*$.
Now $G$ is a {\em 2-sum}
of $G\p$ and $G^*$, i.e $G\p$ and $G^*$ hinged together at an edge, so for any $\delta$ 
$(G,\delta)$ has a
realization if and only if $(G^*, \delta)$ has a realization and
$(G', \delta)$ has realization. Furthermore, either
$\Phi_f^2(G,\delta) = \Phi_f^2(G^*,\delta)$ or 
$\Phi_f^2(G,\delta) = \emptyset$. Note this property holds
if we add more vertices to $G'$ by Henneberg-I steps.
Thus, we have 
$\Phi_f^2(G_m,\delta)= \Phi_f^2(G_{m-1},\delta)$ or
$\Phi_f^2(G_m,\delta)= \emptyset$.
\end{proof}

\subsection{Characterizing Simple 1-dof Henneberg-I
graphs with efficient configuration spaces}

For Simple 1-dof Henneberg-I graphs $G$,
a natural choice of configuration space parameter 
is its base non-edge.
We simply adopt this choice of parameter since it 
guarantees a complete configuration space of 
low realization complexity, i.e., quadratically solvable in 
time linear in $|V|$,
as mentioned in the introduction.  Unlike general Tree- or Triangle-
decomposable graphs, 
since Henneberg-I graphs have a single base edge, they
are some times called {\em ruler and compass constructible or RCC graphs}).

Note that this realization process could lead to an exponential 
combinatorial explosion because there 
are 2 possible orientations for each point $p(v)$ and only one of 
them may successfully lead to a realization of the entire EDCS.
However, we will show in Observation \ref{obs:linear-realization-general} 
that we can circumvent this problem by encoding 
along with each parametrized configuration $\delta^*(f)$,
one (or all) of the {\em orientations}  
$\sigma$ (defined below) of its corresponding realizations. 
Thus the realization complexity is essentially linear in $|V|$.

With this in mind, we only need to characterize which 
Simple 1-dof Henneberg-I 
graphs $G$ have low sampling complexity 
for their configuration space on the base non-edge $f$.
Specifically, this is a 1-parameter
configuration space,  and hence it consists of a union of intervals. 
The sampling complexity is thus 
the complexity of determining the endpoints of these intervals,
starting with $(G,\delta)$ as input. 

In order to quantify and define {\em low sampling complexity}
we prove a crucial result Theorem \ref{thm:config-space} 
that gives a {\sl combinatorial meaning} to the endpoints of the intervals 
in the configuration space $\Phi_f^2(G,\delta)$.
The theorem relies on a technical Lemma
\ref{lem:algebraic}
that gives combinatorial description of the configuration space.
The proof requires basic algebra and real analysis.

\subsubsection{Combinatorial meaning of configuration space boundary}

We first formally define the {\em orientation} of a realization
of a Henneberg-I graph.
As mentioned above, given an EDCS $(H,\delta)$ where $H$ is a Henneberg-I graph
with base edge $f$,
for each Henneberg-I step $v_k \triangleleft (u_k,w_k)$, 
if the coordinates for the point realizations $p(u_k)$ and 
$p(w_j)$ are known and
the values $\delta(v_k,u_k)$ and  $\delta(v_k,w_k)$
are also known, the posssible coordinates for the point $p(v_k)$ can be
determined by a corresponding 
simple ruler and compass algebraic construction (solving a 
quadratic equation in 1 variable).
If the triangle is not a trivial one (three vertices are not
collinear), there are two choices for the coordinates of $p(v_k)$.
We say each of these choices is an {\em orientation} $\sigma_k$ for the
Henneberg-I step $k$. 
If we specify an orientation for each Henneberg-I step in a construction
sequence (partial order) of $H$ from $f$, 
yielding a corresponding sequence (partial order) $\sigma$,
we say 
that a realization of $(H,\delta)$ has an {\em orientation} $(\sigma, f)$. 

\begin{figure}[h]
\psfrag{1}{$v_1$}
\psfrag{2}{$v_2$}
\psfrag{3}{$v_3$}
\psfrag{4}{$v_4$}
\psfrag{5}{$v_5$}
\psfrag{6}{$v_6$}
\psfrag{7}{$v_7$}
\psfrag{8}{$v_8$}
\psfrag{9}{$v_9$}
\psfrag{10}{$v_{10}$}
\begin{center}
\includegraphics[width=12cm]{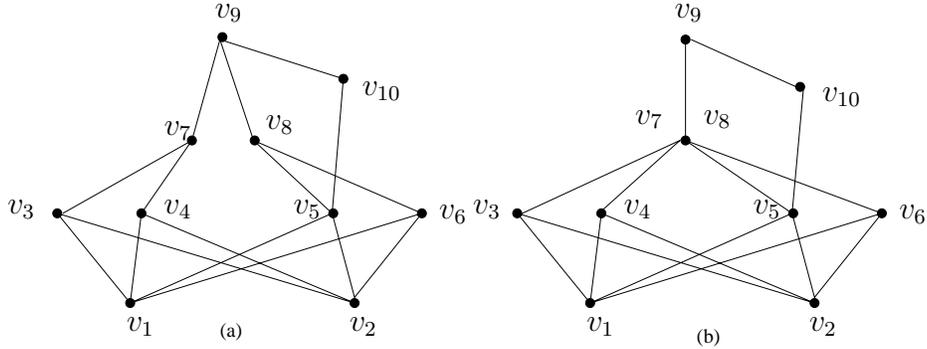}
\end{center}
\caption{When $p(v_7)$ and $p(v_8)$ are coincident, distance $\delta^*(v_5,v_9)$
is not a function of $\delta^*(v_1,v_2)$.}
\label{F:equal-distance}
\end{figure}

In fact, observe that this is  a 1-1 correspondence provided 
$\delta$ assigns distinct distances  to the
edges of $H$. I.e,  for any such $\delta$, 
there exists at most one 2D realization $p$
of $(H, \delta)$,
when an orientation $(\sigma, f)$ is specified.
The coordinates of $p(v_k)$
are not unique
only if at the $k^{th}$ step of the construction sequence $c$ 
the vertex $v_k$  is constructed from 
vertices $u_K$ and $w_k$ for which 
$p(u_k)$ and $p(w_k)$ are coincident and $\delta(v_k,u_k)$ 
is equal to  $\delta(v_k,w_k)$. See Figure~\ref{F:equal-distance}. 


Now consider an EDCS $(G,\delta)$ where 
$G$ is a Simple 1-dof Henneberg-I 
graph with base non-edge $f,$ and assume $\delta$ assigns distinct values.
For any such $\delta$ and 
distance assignment $\delta^*(f)$ distinct from the values
assigned by $\delta$,
an orientation $(\sigma, f)$ (and realization) for $(G\cup f, \delta, 
\delta^*)$
gives a corresponding orientation (and realization) for $(G, \delta)$.
At any construction step, we can regard
$\delta^*(u, w)$ for the base pair of vertices  
as a function of $\delta^*(f)$. The next lemma analyzes this function
to give a combinatorial description for $\Phi_{f}^2(G, \delta)$.

\begin{lemma} 
\label{lem:algebraic}
Given an EDCS $(G, \delta)$ where $G$ is a 
Simple 1-dof Henneberg-I graph with base non-edge $f = (v_1,v_2)$, if (1)
for all Henneberg-1 steps $v \triangleleft (u, w)$ in the construction sequence 
starting from $f$, the two edge distances $\delta(v, u)$ and
$\delta(v, w)$ are distinct
 and (2) an orientation $\sigma$ is specified for the Henneberg-I construction sequence 
starting from $f$, then the following hold:

\begin{enumerate}

\item $\Phi_{f}^2(G,\delta)$ is a set
of closed real intervals or empty;

\item For any interval endpoint $\delta^*(f)$ in $\Phi_{f}^2(G,\delta)$, there is a
unique realization for $(G \cup f, \delta, \delta^*(f))$ with the
orientation $\sigma$ and there exists a Henneberg-I step 
$v \triangleleft (u, w)$ such that the three vertices $v$, $u$
and $w$ are collinear in this unique realization;

\item For any pair of vertices $(u,w)$  and any realization $p$ of 
$(G\cup f),\delta, \delta^*(f))$
the distance $\delta^*_p(u,w)$ is a continuous function of
$\delta^*(f)$ on each closed interval of $\Phi_{f}^2(G,\delta)$. Furthermore, 
for any vertex, $v$, the
coordinates of the point $p(v)$ 
 are continuous functions of
$\delta^*(f)$ on each closed
interval of $\Phi_f^2$,  if we pin the coordinates of 
 $p(v_1)$ to be $(0,0)$
and the $y$-coordinate of $p(v_2)$ to be 0.

\end{enumerate}
\end{lemma}

The proof of this lemma involves basic algebra and real analysis. 
The idea is to do a ruler-and-compass realization sequence that follows 
a Henneberg-I construction sequence and check how each
Henneberg-I step will change the configuration space on the base
non-edge.
In the following, we will loosely use ``Henneberg construction sequence'' also
to refer to the corresponding ruler-and-compass realization sequence.

\medskip
\noindent
\begin{proof} {\bf [Lemma \ref{lem:algebraic}]}
We prove by induction on the length of the given Henneberg-I construction sequence 
starting from $f$. 

In the base case, the length of the given Henneberg-I construction sequence 
is 1. Suppose $v_3$ is the only other vertex. 
By the triangle inequality,
we know $\Phi_{f}^2(G,\delta)$ 
is $[|\delta(v_3,v_1) - \delta(v_3,v_2)|, |\delta(v_3,v_1)+\delta(v_3,v_2)|]$,
 so (1) and (2) are satisfied. For (3), we only need to
consider whether the coordinates of $p(v_3)$ which we denote as 
 $(x_{v_3},y_{v_3})$ are a continuous function of
$\delta^*(f)$. Denote $R_{1}=\delta(v_1,v_3)$,
$R_{2}=\delta(v_3,v_2)$ and $R_{3}=\delta^*(v_1,v_2)= \delta^*(f)$. 
We can compute 
\begin{equation}
x_{v_3}=\frac{ R_1^2 +R_{3}^2 -R_{2}^2}
{2R_{3}}
\end{equation}
\begin{equation}
y_{v_3}=\frac{\sqrt
{(R_{1}+R_{2}+R_{3})
(R_{1}+R_{2}-R_{3})
( R_{1}-R_{2}+R_{3})
(-R_{1}+R_{2}+R_{3})}}
{2R_{3}}.
\end{equation}
Note that since $R_{3}$ is not 0, both $x_{v_3}$ and $y_{v_3}$ are
continuous fuctions of $R_{3}$, which is our $\delta^*(f)$ now.  

By induction hypothesis, we assume that (1), (2) and (3) hold
for a Simple 1-dof Henneberg-I graph $G_{k-1} = (V,E)$ with base non-edge $f$
with less than $k$ Henneberg steps.
Suppose we get a new graph  $G_k$ by one more Henneberg-I step
$v_k \triangleleft (u_k, w_k)$ with base vertices 
$u_k,w_k$ in $G_{k-1}$. I.e.,  $G_k = (V \cup v_k,E
\cup (v_k,u_k) \cup (v_k,w_k))$. We will prove (1),
(2) and (3) hold for $G_k$.

\begin{figure}[h]
\begin{center}
\includegraphics[width=7.5cm]{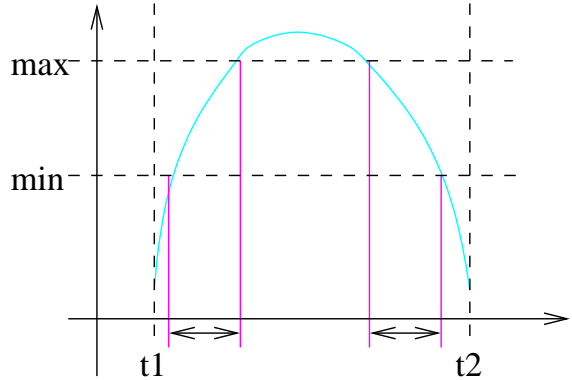}
\end{center}
\caption{For Lemma \ref{lem:algebraic}. 
New constraint on $\delta^*(u_k, w_k)$ changes the interval endpoints in $\Phi_f^2(G_k,\delta).$}
\label{F:RccStepN}
\end{figure}

According to the Statement (3) of the induction hypothesis, 
in the realization $p$ of $G_{k-1}$ with a fixed orientation
$\sigma$, for any pair of vertices $(u,w)$ of $G_{k-1}$, the distance value 
$\delta^*_p(u,w)$ is a continuous function, say $p_{u,w}$,  of $\delta^*(f)$. 
We extend the realization $p$ to the 
newly added vertex $v_k$. Now the edges
$(v_k,u_k)$ and $(v_k,w_k)$ will restrict 
$\delta^*_p(u_k,w_k)$ to be 
in $[min, max]$ where $min = |\delta(v_k,u_k) - \delta(v_k,w_k)|$ 
and $max=\delta(v_k,u_k) + \delta(v_k,w_k)$. 
This restriction will create
new candidate interval endpoints in 
$\Phi_{f}^2(G_k,\delta)$, namely 
$p_{u_k,w_k}^{-1}(\delta^*_p(u,v)), y \in 
[|\delta(v_k,u_k) - \delta(v_k,w_k)|,|\delta(v_k,u_k) + \delta(v_k,w_k)]$, 
as is shown in
Figure~\ref{F:RccStepN}. Since these new candidate interval 
endpoints in 
$\Phi_{f}(G_k,\delta)$ 
correspond to the realization in which $p(u_k)$, $p(v_k)$ and
$p(u_k)$ are collinear, (1) and (2) are also true for graph
$G_k$.

To show the induction step for (3), take any non-edge $(u,w)$. 
We have:

\begin{equation}
\label{distance}
\delta_p^*(u,w)=\sqrt{(x_u-x_w)^2+(y_u-y_w)^2}
\end{equation}

If $u \not= v_k$ and $w \not= v_k$, $\delta^*_p(u,w)$ is clearly a
continuous function of $\delta^*(f)$, so we only need consider the case that
either $u = v_k$ or $w = v_k$.

For convenience, first rotate and translate 
the coordinate system so that in the triangle $\triangle(u_k,w_k,v_k)$, 
$u_k$ is at the origin and
$\vec{u_k,w_k}$ is the $x$-axis. 
Without loss of generality, let $p(v_k)$ be located
above the line joining $p(u_k)$ and $p(w_k)$, by the given orientation $\sigma$ in the statement
of the Lemma.  
Denote $R_{1}=\delta(v_k,u_k)$,
$R_{2}=\delta(v_k,w_k)$ and $R_{3}=\delta_p^*(u_k,w_k)$. Then,
\begin{equation}
x_{v_k}=\frac{R_{1}^2+R_{3}^2-R_{2}^2}{2R_3}
\end{equation}
and
\begin{equation}
y_{v_k}=\frac{\sqrt{(R_{1}+R_{2}+R_{3})(R_{1}+R_{2}-R_{3})(R_{1}-R_{2}+R_{3})(-R_{1}+R_{2}+R_{3})}}{2R_{3}}.
\end{equation}

Since we have restricted $R_{1} \not= R_{2}$, we have $R_{3}>0$. 
Consider the rotation and
translation that now put the point $p(v_1)$ at the origin and  $p(v_2)$
on the $x$-axis as in the statement of the Lemma.
Denote the rotation angle as $\beta$. Then we have:

\begin{equation} 
\cos \beta=\frac{x_{w_k}-x_{u_k}}{R_3} 
\end{equation} 

\begin{equation} 
\sin \beta==\frac{y_{w_k}-y_{u_k}}{R_3}
\end{equation}

So we can get the transformed coordinates of $p(v_k)$:

\begin{equation}
x_{v_k}^{*}=x_{u_k}+x_{w_k}*\cos \beta+y_k*\sin \beta
\end{equation}
\begin{equation}
y_{v_k}^{*}=y_{u_k}+x_{w_k}*\sin \beta+y_k*\cos \beta
\end{equation}

$x_{v_k}$ and $y_{v_k}$ is a function of $x_{u_k}$, $y_{u_k}$,
$x_{w_k}$ and $y_{w_k}$,
so for any value $\delta^*(f)$ 
over a closed interval, the coordinates $p(v_k)$ can be 
expressed as a function of $\delta^*(f)$ using radicals. 
So, $\delta(u,w)$ in
equation~\ref{distance} is a continuous function of $\delta^*(f)$ 
even if $u = v_k$ or
$ w = v_k$ and this proves the induction step of Statement (3) of the 
Lemma~\ref{lem:algebraic} for  
graph $G_{k}$. 
\end{proof}

\medskip
{\bf Remark.}
In Lemma~\ref{lem:algebraic}, we require that 
the two distances
$\delta(v_k, u_k)$ and $\delta(v_k, w_k)$ are not equal
for the $k^{th}$
Henneberg-I step $v_k \triangleleft (u_k,w_k)$. This requirement guarantees that
the two points $p(u_k)$
and $p(w_k)$ in a realization $p$ for $(G_{k-1},\delta)$ are not 
coincident, whereby the quantity $R_3 > 0$  and thus we can 
use a continuity argument.

Now we can state the theorem that interests a combinatorial meaning to the 
configuration space of a Simple 1-dof Henneberg-I graph using the notion
of {\em extreme graphs} defined earlier.

\begin{theorem}
\label{thm:config-space}
Given an EDCS $(G,\delta)$ where $G$ is a Simple 1-dof Henneberg
graph with a base non-edge $f$,
the endpoints of the intervals in the 
configuration space $\Phi_f^2(G,\delta)$ are contained in the set:
${\cal E}(G,\delta) :=
\bigcup\limits_\sigma\bigcup\limits_{1\le k\le |V|-2} \bigcup\limits_{1\le j\le 2}
\{\delta^{X_{k,j}^{1,\sigma}}(f), \delta^{X_{k,j}^{2,\sigma}}(f),\ldots \}$; 
\\
where  
$\delta^{X_{k,j}^{m,\sigma}}(f)$
denotes the length or distance value of $f$ in the $m^{th}$ 
realization $p_m$ with orientation sequence $\sigma$
of the  $k^{th}$ {\em extreme  EDCS} $X_{k,j}$ determined by the pair $(G,f)$.
\end{theorem}

Fact~\ref{fact:extremeWell}(1) guarantees that the graph 
corresponding to this extreme EDCS is wellconstrained provided   
the two vertices incident on the new edge 
were not previously in a wellconstrained subgraph. 
If they were in a wellconstrained subgraph,
then 
the corresponding two EDCSs $X_{k,1}$ and $X_{k,2}$
can be left undefined, and the corresponding interval endpoints do not
appear in ${\cal E}(G,\delta)$ by Fact~\ref{fact:extremeWell}(2).

\noindent
\begin{proof}
The proof directly follows from Statement (2) of Lemma~\ref{lem:algebraic} (2).
\end{proof}

\begin{figure}[h]
\psfrag{xx}{$\delta^*(f)$}
\psfrag{y1}{$p_{u_k,v_k}(\delta^*(f))$}
\psfrag{y2}{$=$}
\psfrag{y3}{$\delta_p^*(u_k,w_k)$}
\psfrag{aa}{$\diamond$}
\psfrag{bb}{$\bullet$}
\psfrag{cc}{$\circ$}
\begin{center}
\includegraphics[width=14cm]{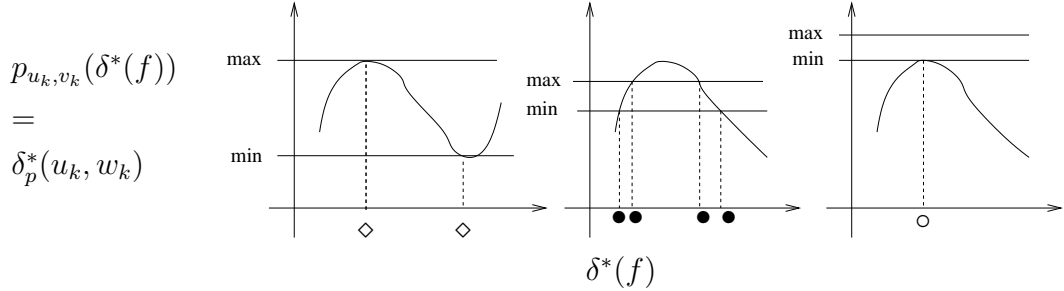}
\end{center}
\caption{
For Observation~\ref{obs:linear-realization-general}.
(Left) shows extreme EDCS configurations 
$\diamond$ 
in ${\cal E}_\sigma(G,\delta)$ 
that are in some proper interval of ${\cal I}_\sigma$, but not endpoints;
(Middle) shows extreme EDCS configurations $\bullet$ that are endpoints
of intervals in  
of ${\cal I}_\sigma$; and (Right) shows extreme EDCS configurations $\circ$
that are isolated points in 
${\cal I}_\sigma$.
}
\label{F:endpoints}
\end{figure}

\begin{observation}
\label{obs:linear-realization-general}
Theorem \ref{thm:config-space} implies 
linear realization complexity of the configuration
space $\phi_f^2(G,\delta)$:
for each candidate orientation sequence $\sigma$,
we can read off a set of intervals ${\cal I}_\sigma$ from 
the description ${\cal E}(G,\delta)$ 
as in Theorem \ref{thm:config-space},
such that a configuration $\delta^*(f) \in {\cal I}_\sigma$
is guaranteed to correspond to a 
realization with the orientation $\sigma$.
Knowing a realizable orientation $\sigma$
for the configuration $\delta^*(f)$
eliminates the combinatorial explosion during a 
linear time ruler-and-compass
realization of $(G \cup f, \delta, \delta^*)$.
\end{observation}

\noindent
\begin{proof}
By Theorem~\ref{thm:config-space} the endpoints of $\Phi_f^2(G,\delta)$ form 
a subset of the candidate set ${\cal E}(G,\delta)$, which we view
as a union over candidate sets for each orientation $\sigma$: 
$\bigcup\limits_\sigma {\cal E}_\sigma(G,\delta)$. 
While every such candidate configuration 
$\delta^*(f)$ is a configuration of an extreme   
EDCS of $G$, not every candidate  configuration
is actually an interval endpoint for $\Phi_f^2(G,\delta)$, nor even an endpoint
of the set of intervals ${\cal I}_\sigma$ required in the statement of the 
Observation.  To see this, 
recall the proof for Lemma~\ref{lem:algebraic}(Figure~\ref{F:RccStepN}); let
$v_k$ be the vertex constructed in the $k^{th}$ step
of the Henneberg construction of $G\cup f$ starting from $f$, and let
$u_k$ and $w_k$ be the base vertices of this step. 
Consider the continuous function $p_{u_k,w_k}$ in the variable $\delta^*(f)$
which gives the distance between $u_k$ and $w_k$ in a particular
realization $p$ with orientation $\sigma$; 
i.e, the value of this continuous function $p_{u_k,v_k}$ evaluated at 
$\delta^*(f)$ is the distance  $\delta^*_p(u_k,w_k)$.
Figure~\ref{F:endpoints} shows that 
based on this continuous function, the 
2 distance values $min=|\delta(v_(k),u_k) - \delta(v_(k),w_k)|$ and 
$max=|\delta(v_(k),u_k) - \delta(v_k,w_k)|$,
all the following four cases are possible for a candidate configuration 
$\delta^{X_{k,j}^{m,\sigma}}(f)$:
neither the left nor the right neighborhood falls into
$\Phi_f^2(G,\delta)$; both the left and the right neighborhood fall into
$\Phi_f^2(G,\delta)$;  the left falls into
$\Phi_f^2(G,\delta)$ but the right does not; and symmetrically
 the right falls into
$\Phi_f^2(G,\delta)$ but the left does not.
In the first case, the candidate configuration is an isolated point
in ${\cal I}_\sigma$. In the second, it is  not an endpoint of any
interval in 
${\cal I}_\sigma$. 
In the third and the fourth cases, it is 
actually an endpoint of an interval in
${\cal I}_\sigma$. 

In other words, in order to
produce such a set of intervals ${\cal I}_\sigma$ from 
the candidate configurations ${\cal E}_\sigma(G,\delta)$, 
we need to check 
whether the left and/or right neighborhood of each such candidate configuration
also belongs to $\Phi_f^2(G,\delta)$, i.e, whether it has
a realization. 
To find out which of the above 4 cases applies, one can check  
if there is any realization with orientation $\sigma$,
for values of $\delta^*(f)$ that are to the left 
(resp. right) of 
$\delta^{X_{k,j}^{m,\sigma}}(f)$,
but before the candidate configuration in ${\cal E}_\sigma(G,\delta)$ that is 
immediately preceding (resp. immediately succeeding)
$\delta^{X_{k,j}^{m,\sigma}}(f)$.
This is straightforward after sorting the set 
${\cal E}_\sigma(G,\delta)$. 
 Since the orientation 
is fixed, checking if such realizations exist 
can be done in linear time with a ruler and compass
construction. 
\end{proof}

\medskip
Based on such a description of the configuration space $\Phi_f^2(G,\delta)$, we say it  has
{\em low sampling complexity} if all of the 
extreme EDCS are 
Tree- or Triangle-decomposable, 
which ensures that the interval endpoints 
$\delta^{X_{i,j}^k}(f)$
in the above theorem 
can be computed essentially using
a sequence of solving one quadratic equation at a time.
This ensures linear complexity in $|V|$
It has additionally been conjectured
these graphs {\sl exactly} capture
Quadratic Solvability and the conjecture
has been proven for planar graphs \cite{bib:Owen02}.

\subsubsection{Forbidden minor characterization for 1-path triangle-free Simple 1-dof Henneberg-I graphs}

\medskip
The next theorem gives a surprising and  exact forbidden-minor characterization 
of a large class of Simple 1-dof Henneberg-I graphs $G$ with
base non-edge $f$ such that for all 
distance assignments $\delta$, 
the EDCS $(G,\delta)$, the  
a configuration space 
$\Phi_f^2(G,\delta)$ has low sampling complexity.
In other words, {\em all} the extreme graphs obtained from $(G,f)$ 
are Tree- or Triangle- decomposable.

\begin{figure}[h]
\psfrag{1}{$v_1$}
\psfrag{2}{$v_2$}
\psfrag{3}{$v_3$}
\psfrag{4}{$v_4$}
\psfrag{5}{$v_5$}
\psfrag{6}{$v_6$}
\psfrag{7}{$v_7$}
\begin{center}  
\includegraphics[width=12cm]{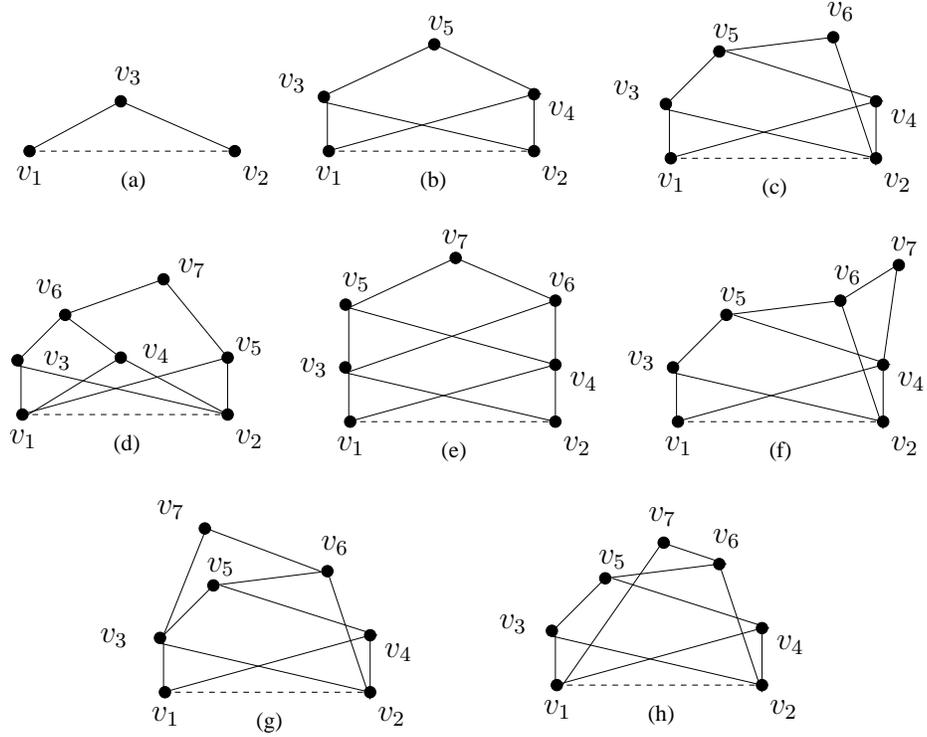}
\end{center}  
\caption{All the 1-path triangle-free Simple 1-dof Henneberg-I graphs with less then 8 
vertices; neither (d) nor (h) has low sampling complexity on base non-edge 
$(v_1,v_2)$ while all the other have; both (d) and (h) have a $K_{3,3}$ minor
while all the others do not have.}
\label{F:triangle-free-basecases}
\end{figure}

A Simple 1-dof Henneberg-I graph with base non-edge $f$ has the {\em 1-path} property
if exactly one vertex other than the endpoints of $f$ has degree 2. 
We say a graph $G$ is {\em triangle-free} if $G$ has no subgraph that is a triangle
(see Figure~\ref{F:triangle-free-basecases}).

\begin{theorem} \label{the:TriangleFreeCase} 
Let $G$ be a triangle-free 1-path 1-dof Henneberg-I graph that represents the
construction path of
$v_{n}$
from base non-edge $f$. Then
\begin{enumerate}
\item
$G$ has 
a configuration space of 
low sampling complexity
if and only if $G$ has no $K_{3,3}$ or $C_{3} \times
C_{2}$ minor;
\item
$G$ has a configuration space of low sampling complexity 
if and only if for any Henneberg-I step $v\triangleleft (u, w)$
associated to $G$ and the base non-edge 
$f$, the (extreme) graph $G \cup (u,w)$ is
a Henneberg-I graph with base edge $(u,w)$.
\end{enumerate}
\end{theorem}

The proof of the theorem relies on several lemmas.

\begin{lemma}
\label{lem:4lemmas}
\begin{enumerate}
\item 
Let $G$ be a 1-path Simple 1-dof Henneberg-I graph with base non-edge $f =(v_1,v_2)$.
Then
\begin{itemize}
\item [1.a]  
 if the number of vertices 
directly constructed with $v_1$ and $v_2$ as base vertices is 3 or more, then $G$ has 
a $K_{3,3}$ minor.
\item [1.b] 
if the number of vertices directly constructed with $v_1$ and $v_2$ as base vertices 
is exactly 2 and 
both $deg(v_1)$ and $deg(v_{2})$ are at least 3, then $G$ has 
a $K_{3,3}$ or $C_{3} \times C_{2}$ minor.
\end{itemize}
\item 
Let $G$ be a 1-path Simple 1-dof Henneberg-I graph with base non-edge $f$. Then
$G$ does not have low sampling complexity on $f$ if either of the following
holds
\begin{itemize}
\item [2.a] 
the number of vertices 
directly constructed with $v_1$ and $v_2$ as base vertices is 3 or more, then $G$ does not
have low sampling complexity on $f$.
\item [2.b] 
the number of vertices directly constructed with $v_1$ and $v_2$ as base vertices 
is exactly 2 and 
both $deg(v_1)$ and $deg(v_{2})$ are at least 3,
then $G$ does not
have low sampling complexity on $f$.
\end{itemize}
\end{enumerate}
\end{lemma}

\noindent
\begin{proof}
Since $G$ is a 1-path Simple 1-dof Henneberg-I graph 
with base non-edge $f$, we use $v_n$ to denote
the last vertex in the construction sequence starting
from $f$. Additionally we use $u_i$( $i = 1,\cdots, m$) to denote the vertices 
constructed with $v_1$ and $v_2$ as base vertices.
As the last vertex in the Henneberg-I sequence, 
$v_n$ has to be different from $v_1$, $v_2$ and all the $u_i$( $i = 1,\cdots, m$)
when $m$ is greater than 1.

\medskip
\noindent
{\bf [1.a]}
We contract all the edges that have at least 1 vertex other than 
$v_1$, $v_2$, $u_i$( $i = 1,\cdots, m$) and $v_n$ (see Figure~\ref{fig:4lemmas}(a)). Since $v_n$
will be adjacent to all $u_i$( $i = 1,\cdots, m$) in the contracted
graph, the contracted graph has a $K_{3,3}$ minor which is induced
 by  $v_1$, $v_2$, $v_n$,
$u_1$, $u_2$ and $u_3$ ($v_1$, $v_2$ and $v_n$
are as one partition and $u_1$, $u_2$ and $u_3$ as the other.

\medskip
\noindent
{\bf [1.b]} 
Consider the possible ways we construct the fifth vertex
following $v_1$, $v_2$, $u_1$ and $u_2$. We denote the fifth vertex
by $v_5$. Because $m=2$, $v_5$ cannot be constructed with
$v_1$ and $v_2$ as base vertices. So, either $v_5$ is constructed with
$u_1$ and $u_2$ as base vertices (see Figure~\ref{fig:4lemmas}(c)) or 
using a base edge whose vertices are
among $v_1$, $v_2$, $u_1$ and $u_2$. For the latter case, 
without loss of generality, we assume $v_5$ is constructed with $v_1$ and $u_1$ (see Figure~\ref{fig:4lemmas}(b)).
For both cases, we contract all the edges which have at least one vertex other than
$v_1$, $v_2$, $u_1$, $u_2$, $v_5$ and $v_n$. 
For the former case shown in Figure~\ref{fig:4lemmas}(b),
there is a $C_3 \times C_2$ minor in the contracted graph where the two triangles
are $\triangle (v_1,u_1,v_5)$ and $\triangle (v_2,u_2,v_n)$; 
for the latter case shown in Figure~\ref{fig:4lemmas}(c),
there is a $K_{3,3}$ minor in the contracted graph where 
$v_1$, $v_2$ and $v_5$
are in one partition and $u_1$, $u_2$ and $v_n$ are in the other.

\begin{figure}[h]
\psfrag{v1}{$v_1$}
\psfrag{v2}{$v_2$}
\psfrag{u1}{$u_1$}
\psfrag{u2}{$u_2$}
\psfrag{u3}{$u_3$}
\psfrag{v5}{$v_5$}
\psfrag{vn}{$v_n$}
\psfrag{(a)}{(a)}
\psfrag{(b)}{(b)}
\psfrag{(c)}{(c)} 
\begin{center} 
\includegraphics[width=10cm]{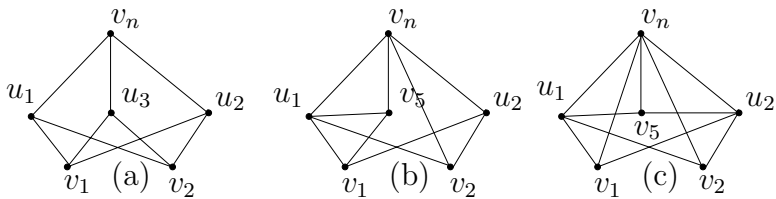}
\end{center}  
\caption{Edge contractions and graph minors for Lemma~\ref{lem:4lemmas}.}
\label{fig:4lemmas}
\end{figure}

\medskip
\noindent
{\bf [2.a]} Assume that $v_n$ is constructed with
$u$ and $w$ as base vertices. 
We will prove by contradiction that the extreme graph $G \cup (u,w)$ 
is not Triangle-decomposable. 
Assume that $G \cup (u,w)$ 
has a Triangle decomposition into clusters
$C_{1}$, $C_{2}$ and $C_{3}$. By the definition of Triangle decomposition,
$C_{1}$, $C_{2}$ and $C_{3}$ are also Triangle-decomposable and wellconstrained. 
We know by Fact~\ref{fact:noWellOn12} that  $v_1$
and $v_2$ cannot both belong to any wellconstrained subgraph, so 
 $v_1$
and $v_2$ cannot both belong to any of $C_{1}$, $C_{2}$ and $C_{3}$.
Vertex $u_1$ is adjacent to both $v_1$ and $v_2$, which are not both in
a cluster, so $u_1$ must be a vertex shared by two different clusters of 
$C_{1}$, $C_{2}$ and $C_{3}$. Similarly, $u_2$
and $u_3$ are shared vertices. Now $u_1$, $u_2$ and $u_3$ 
are the three shared vertices (Refer to Figure~\ref{F:triangleDecomposition}) 
but $v_1$ and $v_2$ are adjacent to
all these three shared vertices which is impossible(see Figure~\ref{F:triangleDecomposition}).   

\medskip
\noindent
{\bf [2.b]}. 
We prove this by contradiction.
Assume that the extreme graph $G \cup (u,w)$ 
is Triangle-Decomposable.
Since $v_n$ has degree 2, $(G\setminus \{v_n\}) \cup (u,w)$
and $G\setminus \{v_n\}$ have the same Triangle-Decomposability,
so $(G\setminus \{v_n\}) \cup (u,w)$ is also Triangle-Decomposable.
Suppose  $(G\setminus \{v_n\}) \cup (u,w)$
has a triangle 
decomposition $C_{1}$, $C_{2}$
and $C_{3}$. 
Observe that $v_1$ and $v_2$ cannot both
belong in the same one of $C_{1}$, $C_{2}$
or $C_{3}$. Otherwise, suppose both $v_1$ and $v_2$ are in
$C_1$.
By Fact~\ref{fact:noWellOn12}, 
if $C_1$ does not contain edge $(u, w)$, $C_1$ will not be wellconstrained,
so $C_1$ must contain edge $(u, w)$. Because $G$ is 1-path, vertices
$u$ and $w$ are the two base vertices of the last constructed vertex $v_n$,
and $C_1$ contains $v_1$ and $v_2$ which are the two vertices of the base
non-edge,
$C_1$ must be the entire graph
 $(G\setminus \{v_n\}) \cup (u,w)$ and this makes $C_2$ and $C_3$ impossible,
 so $v_1$ and $v_2$ do not both lie in any one of $C_{1}$, $C_{2}$
and $C_{3}$. This fact together with the fact that $u_1$ is adjacent to both $v_1$ and $v_2$
implies that $u_1$ has to be a shared vertex for the Triangle-decomposition.
Similarly, $u_2$ also has to be a shared vertex for the Triangle-decomposition.
Two shared vertices always belong in a same Triangle-decomposition component, so
without loss of generality suppose $u_1$ and $u_2$ are both in $C_1$. Now $v_1$
and $v_2$ are both adjacent to $u_1$ and $u_2$ and $v_1$
and $v_2$ are not in the same Triangle-decomposition component. This implies
one of $v_1$ is in $C_1$ as a non-shared vertex and the other is 
the third shared vertex for the Triangle-decomposition.
See Figure~\ref{F:four-lemma-2b}.

\begin{figure}[h] 
\begin{center}
\psfrag{v1}{$v_1$}
\psfrag{v2}{$v_2$}
\psfrag{u1}{$u_1$}
\psfrag{u2}{$u_2$}
\psfrag{u}{$u$}
\psfrag{w}{$w$}
\psfrag{vn}{$v_n$}
\psfrag{(a)}{(a)}  
\includegraphics[width=5cm]{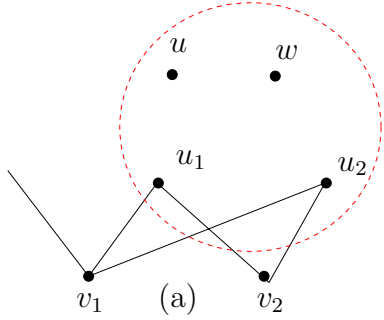}
\end{center}  
\caption{Proof of Lemma~\ref{lem:4lemmas} (2b).}
\label{F:four-lemma-2b}
\end{figure} 
 
Next it is not hard to show that  
$G$ is also a Simple 1-dof Henneberg-I graph with base non-edge
$(u_1,u_2)$. 
By Fact~\ref{fact:noWellOn12}, no subgraph of $G$ containing
both vertices of a base non-edge is well-constrained. Now $C_1$
contains both $u_1$ and $u_2$ which are the two vertices of a base
non-edge for $G$, so for $C_1$ to be wellconstrained, the edge $(u, w)$
has to belong in $C_1$. This implies that both $C_2$ and $C_3$ are 
subgraphs of $G$.
By Fact~\ref{fact:wellcondition}, a well-constrained subgraph of
a 1-dof Henneberg-I graph has to be a Henneberg-I graph, so 
$C_2$ is a Henneberg-I graph. Further, according to the 
order of vertices in the Henneberg-I construction sequence of 
$G$ starting from $(v_1, v_2)$ and the conclusions of the previous
paragraph, the edges $(u_1, v_2)$ 
and $(u_2, v_2)$ have to be the
base edges for Henneberg-I graphs $C_2$ and $C_3$ respectively.
This restricts $C_2$ and $C_3$ to be pure edges, otherwise, a vertex in $C_2$ (resp $C_3$)
other than $u_2$ (resp. $u_1$) and $v_2$ has degree of 2 and this contradicts to the
1-path property of $G$( the only vertex of $G$ with degree of 2 is in
$C_1$). Both $C_2$ and $C_3$ are pure edges so $deg(v_2)$ is 2. This contradicts 
to the both $deg(v_1)$ and $deg(v_2)$ are at least 3. 
\end{proof}





\begin{lemma}
\label{lem:equivalence}
\begin{enumerate}
\item 
Given a 1-dof Henneberg-I graph $G$
with base non-edge $f=(v_1,v_2)$, if $u_{1}$ and $u_{2}$ are the
only vertices constructed with $v_1$ and $v_2$ as base vertices and 
$deg(v_1)$ is 2, then 
\begin{itemize}
\item
(1.a) $G \setminus \{v_1\}$ is a simple 1-dof Henneberg-I graph
with base non-edge $(u_1,u_2)$;
\item
(1.b) $G \setminus \{v_1\}$ 
has low sampling complexity on $(u_1,u_2)$ if and only if 
$G$  has low sampling complexity on $f$. 
\end{itemize}
\item
Given a 1-dof Henneberg-I graph $G$
with base non-edge $f=(v_1,v_2)$, if $u_{1}$ and $u_{2}$ are the
only vertices constructed with $v_1$ and $v_2$ as base vertices and 
both $deg(v_1)$ and $deg(v_2)$ are 2, then 
\begin{itemize}
\item
(2.a) $G \setminus \{v_1,v_2\}$ is a simple 1-dof Henneberg-I graph
with base non-edge $(u_1,u_2)$
\item
(2.b) $G \setminus \{v_1,v_2\}$
has low sampling complexity on $(u_1,u_2)$ if and only if 
$G$  has low sampling complexity on $f$. 
\end{itemize}
\end{enumerate}
\end{lemma}

\begin{figure}[h]
\begin{center}
\psfrag{v1}{$v_1$}
\psfrag{v2}{$v_2$}
\psfrag{u1}{$u_1$}
\psfrag{u2}{$u_2$}
\psfrag{(a)}{(a)}
\psfrag{(b)}{(b)}
\includegraphics[width=8cm]{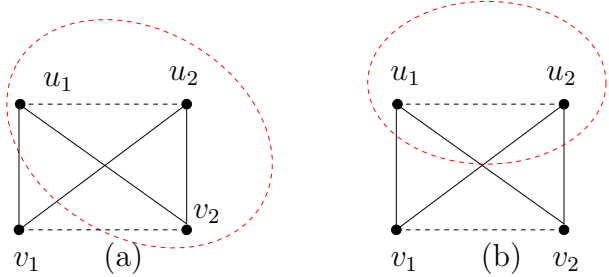}
\end{center}  
\caption{Proof of Lemma~\ref{lem:4lemmas} (2b).}
\label{F:lemma-equivalence}
\end{figure}

\noindent
\begin{proof}
\noindent
{\bf [1.a]} $G$ is a 1-dof Henneberg-I graph $G$
with base non-edge $f$, so we have a Henneberg-I sequence $s_1$ starting
from $f$. Now $u_{1}$ and $u_{2}$ are the
only vertices constructed with $v_1$ and $v_2$ as base vertices,
so the $u_1 \triangleleft (v_1,v_2)$ and  $u_2 \triangleleft (v_1,v_2)$ 
are the first two Henneberg-I steps. 
We can modify $s_1$  to get a new Henneberg-I sequence $s_2$ by starting
from $(u_1, u_2)$ and follow by Henneberg-I
steps $v_1 \triangleleft (u_1,u_2)$ and  $v_2 \triangleleft (u_1,u_2)$.
So,  $(u_1,u_2)$ is a base non-edge for $G$. Since $deg(v_1)$ is 2, 
if we remove $v_1$ from $G$, $(u_1,u_2)$ is still 
a base non-edge for the remaining graph $G \setminus \{v_1\}$, which we
denote by $G\p$. We have proved (1.a).

\noindent
{\bf [1.b]} Recall that a graph $G$ has low sampling complexiy on base non-edge $f$
if and only if all the corresponding extreme graphs are triangle decomposable.
Compare all the extreme graphs corresponding to $G$ (with base non-edge $f$) and 
$G\p$ (with base non-edge $(u_{1},u_{2})$), the former always has $v_1$ as an extra vertex
of degree two. Since adding/removing a vertex and two edges adjacent to the vertex
preserves the Triangle-Decomposability, $G\p$ has low sampling complexity on base non-edge 
$(u_{1},u_{2})$ if and only if
$G$ has low sampling complexity on $f$.

For (2.a) and (2.b), we can directly extend the argument for (1.a) and (1.b) except
that we need add/remove both $v_1$ and $v_2$ that have degree of 2. 
\end{proof}

\medskip
Now we can give the proof of Theorem~\ref{the:TriangleFreeCase}.

\medskip
\noindent
\begin{proof}
{\bf [Theorem~\ref{the:TriangleFreeCase}]}
One direction of (2) in Theorem~\ref{the:TriangleFreeCase}
is trivial:
if the (extreme) graph $G \cup (u_k,w_k)$ is
a Henneberg-I graph with base edge $(u_k,w_k)$
for any Henneberg-I construction $v_k\triangleleft u_k, w_k$ 
associated to $G$ and the base non-edge
$(v_1,v_2)$, then by the defintion of low sampling complexity,
 graph $G$ has low sampling
complexity on $(v_1, v_2)$. 

\medskip
We prove the reverse direction of (1).
Consider the number of vertices which are directly constructed on $(v_1,v_{2})$.
Denote it by $m$.

\noindent
{\bf [Case 1]} If $m$ = 1, $G$ can only be trivially be two edges and $G$ has low
sampling complexity 
on $(v_1,v_{2})$. 

\noindent
{\bf [Case 2]} If $m$ = 3, by Lemma~\ref{lem:4lemmas} (1.a) and 
Lemma~\ref{lem:4lemmas}(2.a), 
$G$ has $K_{3,3}$ or $C_{3}*C_{2}$
minor
and $G$ does not have low complexity 
on $(v_1,v_{2})$ .

\noindent
{\bf [Case 3]} If $m$ = 2 and both $deg(v_1)$ and $deg(v_{2})$ are 3 or more, by
Lemma~\ref{lem:4lemmas} (1.b) and Lemma~\ref{lem:4lemmas} (2.b), $G$ 
does not have low sampling complexity on $(v_1,v_{2})$  and has
$K_{3,3}$ or $C_{3}*C_{2}$
minor.

\noindent
{\bf [Case 4]} If $G$ has low sampling complexity on $(v_1,v_{2})$ and $m$ = 2, 
by Lemma~\ref{lem:4lemmas} (2.b), either $deg(v_1)$ or $deg(v_{2})$ is 
2. Without loss of generality, suppose $deg(v_1)$ is 2 and $u_{1}$, $u_{2}$ are the two
vertices constructed with $v_1$ and $v_2$ as base vertices. 
Denote $G \setminus \{v_1\}$ by $G\p$.

Since $G$ has
low sampling complexity on $(v_1,v_{2})$, by Lemma~\ref{lem:equivalence} (1.a) and (1.b)
so $G\p$ has low sampling complexity on $(u_{1},u_{2})$. 

We can prove now by contradiction that if $G$ does not have low sampling complexity
on $(v_1,v_{2})$, then $G$ has a $K_{3,3}$ or
$C_{3}*C_{2}$ minor. Assume not, then we can find a $G$ with minimum
number of vertices such that $G$ does not have low sampling complexity on $(v_1,v_2)$ 
and $G$ does not have a $K_{3,3}$ or $C_{3}*C_{2}$ minor.  
Consider the number of vertices directly constructed on $v_1$ and $v_2$,
$G$ cannot be in Case 1, Case 2 or Case 3. So, $G$ can only be in Case 4. Since $G$ 
does not have low sampling complexity on $(v_1,v_{2})$, $G\p$ does not 
have low sampling complexity on $(u_{1},u_{2})$. Graph $G$ has no
$K_{3,3}$ or $C_{3}*C_{2}$ minor so $G\p$ does not have $K_{3,3}$ or
$C_{3}*C_{2}$ minor either. Graph $G\p$ has less number of vertices than
$G$ and does not have low sampling complexity on $(u_{1},u_{2})$ 
and it does not have $K_{3,3}$ or $C_{3}*C_{2}$ minor, so we have a
contradiction.

\medskip
For (2) and the reverse direction of (1), we will prove a stronger argument: if Henneberg-I graph $G$ 
has low sampling complexity on base non-edge $(v_1,v_{2})$, then 
all extreme graph $\tilde {G_k} = G \cup (u_k, w_k)$ 
is also a Henneberg-I graph where $u_k$ and $w_k$
are the two base vertices for the $k$'th Henneberg-I step $v_k \triangleleft (u_k, w_k)$. 
We prove this by induction on the number of vertices of $G$. 

Base case: if the number of vertices of $G$ is 3, $G$ has low sampling 
complexity on $(v_1,v_{2})$ and $\tilde {G_3}$ is an edge,
a trivial Henneberg-I graph. 

Assume that $\tilde {G_k}$ is Henneberg-I if $k \leq n$. 
For the induction step, we will prove $\tilde {G_{k+1}}$ is also Henneberg-I.
Recall the above four cases. Since $G$ has low sampling complexity on $(v_1,v_2)$ and
Case 1 is trivial, so we only need to consider Case 4. Now $G$ has low
sampling complexity on $(v_1,v_2)$ implies that $G\p$ 
has low sampling complexity on base non-edge $(u_{1},u_{2})$; and the extreme
graph of $G\p$ is a Henneberg-I graph by assumption, so $\tilde {G}$ is also
Henneberg-I graph. \end{proof}

\bigskip
Next we show that although the low sampling complexity of the graphs characterized in 
Theorem \ref{the:TriangleFreeCase}
have low sampling complexity results from Triangle-decomposable extreme
graphs, their configuration space description (i.e., interval endpoints)
can be obtained using a direct method, {\sl without realizing}
the extreme graphs. 

\begin{observation}
\label{obs:backward}
Given a triangle-free 1-path 1-dof Henneberg-I graph $G = (V, E)$ 
with base non-edge $f$,
if $G$ has low sampling complexity on $f$, then the
configuration space $\delta$, $\Phi^2_f(G,\delta)$
can be computed by an $O(|V|)$ algorithm.
\end{observation}

We use a {\em quadrilateral diagonal interval mapping} by which we mean the 
possible distance intervals of one diagonal $f$ under 4 distance equality constraints $\delta(e_1),\ldots,\delta(e_4)$ 
for four edges $e_1,\ldots,e_4$  of a quadrilateral and a 
distance interval constraint
$[\delta^l(e),\delta^r(e)]$ for the other diagonal $e$ of the quadrilateral. 
The distance intervals for $f$ are obtained from the implicit curve that 
relates the length $\delta^*(e)$ and the length $\delta^*(f)$. This curve is  
can be viewed as equating the volume  of the tetrahedron 
formed by the edges $e_1,\ldots,e_4,e,f$  to zero.  
The curve has the following useful property:
given a value for the length of $e$, say $\delta^l(e)$ (resp. $\delta^r(e)$), there are 
0,1 or 2 distinct 
corresponding values for the length of $f$. In the case that there are 2 values, we denote them
$\delta^l_l(f)$  and $\delta^l_r(f)$ 
(resp. $\delta^r_l(f)$ and $\delta^r_r(f)$).
It is possible that for some value of $\delta^l(e)$ (resp. $\delta^r(e)$) 
the 2 corresponding lengths of $f$ co-incide to 1 value, i.e., 
$\delta^l_l(f)  = \delta^l_r(f)$ 
(resp. $\delta^r_l(f)  = \delta^r_r(f)$. 
This happens for the overall maximum and minimum values for the length of $e$ 
that are permitted by the curve, which are denoted $\delta_{min}(e)$,
$\delta_{max}(e)$.
These values are determined easily by triangle inequalities using the 
2 triangles based on the edges $e_1,\ldots,e_4,e$. 
It is also  possible that 
for some value of $\delta^l(e)$ (resp. $\delta^r(e)$) 
there is {\sl no} corresponding value for the length of $f$.

\begin{figure}[h]
\psfrag{min}{\tiny{$\delta^l(e)$}}
\psfrag{max}{\tiny{$\delta^r(e)$}}
\psfrag{1}{\tiny{$\delta^r_l(f)$}}
\psfrag{2}{\tiny{$\delta^r_r(f)$}}
\psfrag{3}{\tiny{$\delta^l_l(f)$}}
\psfrag{4}{\tiny{$\delta^l_r(f)$}}
\psfrag{5}{\tiny{$\delta_{min}(f)$}}
\psfrag{6}{\tiny{$\delta_{max}(f)$}}
\psfrag{e}{\tiny{$\delta^*(e)$}}
\psfrag{f}{\tiny{$\delta^*(f)$}}

\begin{center}
\includegraphics[width=15cm]{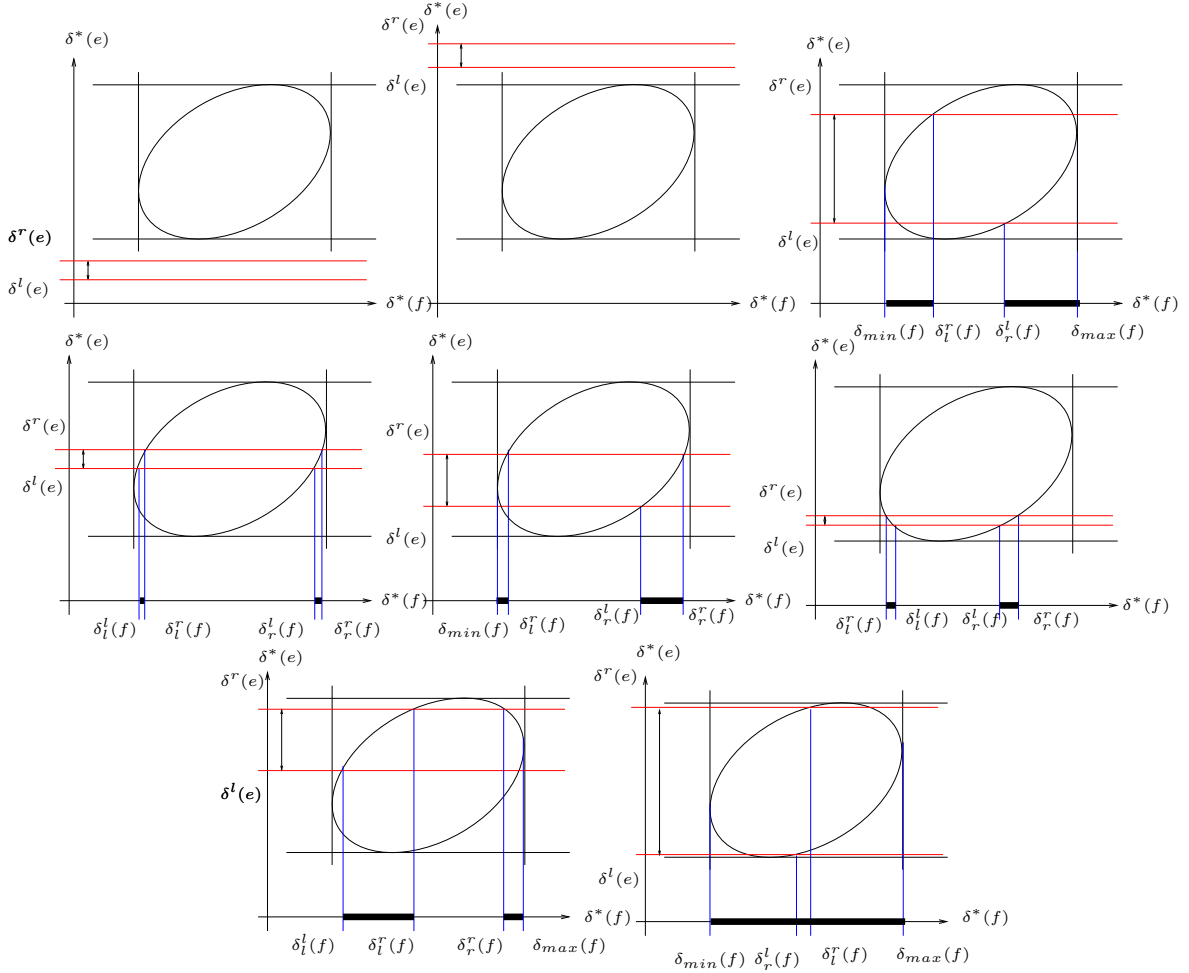}
\end{center}
\caption{
Cases
that must be distinguished in determining the distance interval for $f$,
given the distance interval
$[\delta^l(e),\delta^r(e)]$ for $e$.
}
\label{fig:ellipse}
\end{figure}

See Figure \ref{fig:ellipse} which illustrates the various cases
that must be distinguished in determining the distance interval for $f$,
given the distance interval 
$[\delta^l(e),\delta^r(e)]$ for $e$. 
The various quantities that come into play 
are: (i) $\delta^l(e)$,  $\delta^r(e)$; (ii) the corresponding lengths (if they exist) for $f$ 
$\delta^l_l(f)$,  $\delta^l_r(f)$, and 
$\delta^r_l(f)$, $\delta^r_r(f)$; and moreover 
(iii)
the overall maximum and minimum values for the lengths of $e$ and $f$ that are permitted
by the curve: $\delta_{min}(e)$,
$\delta_{max}(e)$,
$\delta_{min}(f)$, and 
$\delta_{max}(f)$  - as mentioned earlier 
these are determined easily by triangle inequalities using the 
2 triangles based on the edges $e_1,\ldots,e_4$ and $e$ (resp. $f$). 


In particular, for 
Figure~\ref{fig:initial} (right), quadrilateral $(v_1,v_2,v_3,v_4)$ has 
4 distance equality constraints: $\delta(v_1,v_3) = 7$,  $\delta(v_2,v_3) = 7$,
 $\delta(v_1,v_4) = 6$ and  $\delta(v_2,v_4) = 8$. Bounded by triangle inequalities in
$\triangle (v_3,v_4, v_5)$, diagonal $\delta (v_3, v_4)$ has an interval constraint 
as $[4,5]$. By quadrilateral diagonal interval mapping, the possible values of the other diagonal
 $\delta (v_1, v_2)$ is 
$[\frac{1}{8}\sqrt{6214-90\sqrt{17}\sqrt{209}},\frac{1}{8}\sqrt{6214+6\sqrt{17}\sqrt{209}}]$
and
$[\frac{2}{5}\sqrt{565-360\sqrt{2}},\frac{2}{5}\sqrt{565+360\sqrt{2}}]$.


For example, using the following steps  we can get $\Phi^2_{f}(G, \delta)$ ($f=(v_1,v_2)$ )
in Figure~\ref{F:triangle-free-basecases}(e):
\begin{itemize}
\item 
{\bf Step 1:}
Get an interval for $\delta^*(v_5,v_6)$ in $\triangle (v_5,v_6,v_7)$, that is: 
$[|\delta(v_5,v_7) - \delta(v_6,v_7)|,$ \\ 
$ |\delta(v_5,v_7) + \delta(v_6,v_7)|]$;

\item
{\bf Step 2:}
In quadrilateral ($v_3,v_4,v_5,v_6$), get the intervals for $\delta^*(v_3,v_4)$
using the distances $\delta(v_3,v_5)$, $\delta(v_4,v_5)$,
$\delta(v_3,v_6)$, $\delta(v_4,v_6)$ and the interval set $\delta^*(v_5,v_6)$ that is
computed in Step (1);

\item
{\bf Step 3:}
In quadrilateral ($v_1,v_2,v_3,v_4$), get the intervals for $\delta^*(v_1,v_2)$
using the distances $\delta(v_1,v_3)$, $\delta(v_1,v_4)$,
$\delta(v_2,v_3)$, $\delta(v_2,v_4)$ and the interval set $\delta^*(v_3,v_4)$ that is
computed in Step (2); the result will be exactly $\delta_f^2(G,\delta)$.
\end{itemize}

A similar algorithm applies for Figure~\ref{F:triangle-free-basecases}(f).

\begin{itemize}
\item
{\bf Step 1:}
Get one interval for $\delta^*(v_4,v_6)$ in $\triangle(v_4,v_6,v_7)$, that is
$[|\delta(v_6,v_7) - \delta(v_4,v_7)|,$ \\
$|\delta(v_6,v_7) + \delta(v_4,v_7)|]$;

\item
{\bf Step 2:}
In quadrilateral ($v_2,v_4,v_5,v_6$), get the intervals for $\delta^*(v_2,v_5)$
using the distances $\delta(v_2,v_4)$, $\delta(v_2,v_6)$,
$\delta(v_5,v_4)$, $\delta(v_5,v_6)$ and the interval $\delta^*(v_4,v_6)$ that is
computed in Step (1);

\item
{\bf Step 3:}
In quadrilateral ($v_2,v_3,v_4,v_5$), get the intervals for $\delta^*(v_3,v_4)$
using the distances $\delta(v_2,v_3)$, $\delta(v_2,v_4)$,
$\delta(v_3,v_5)$, $\delta(v_4,v_5)$ and the interval set $\delta^*(v_2,v_5)$ that is
computed in Step (2); 

\item
{\bf Step 4:}
In quadrilateral ($v_1,v_2,v_3,v_4$), get the intervals of $\delta^*(v_1,v_2)$
using the distances $\delta(v_1,v_3)$, $\delta(v_1,v_4)$,
$\delta(v_2,v_3)$, $\delta(v_2,v_4)$ and the interval set $\delta^*(v_3,v_4)$ that is
computed in step (3); the result is exactly $\delta_f^2(G,\delta)$.
\end{itemize}

In the Figure~\ref{F:triangle-free-basecases}(e) the two quadrilaterals
for Step ($i$) and step ($i+1$) do not share any edges while
for Figure~\ref{F:triangle-free-basecases}(f) the two quadrilaterals
for Step ($i$) and step ($i+1$) may share two edges. Generally 
the number of the quadrilateral diagonal interval mapping steps is
between $|V|/2$ and $|V|$. Now we give the proof for 
Observation~\ref{obs:backward}.

\medskip
\noindent
\begin{proof}{\bf [Observation~\ref{obs:backward}]}
In fact the observation is subsumed in the proof of Theorem~\ref{the:TriangleFreeCase}.
If the 1-path triangle-free graph $G = (V, E)$ has low sampling complexity
on base non-edge $f=(v_1,v_2)$, we only have three possible cases. {\bf [Case 1]} $|V|$ is 3. 
{\bf [Case 2]} Exactly 2 vertices $v_3$ and $v_4$ are constructed on $(v_1, v_2)$ by
Henneberg-I steps
and both 
$deg(v_1)$ and $deg(v_2)$ are 2. {\bf [Case 3]} Exactly 2 vertices $v_3$ and $v_4$ are
constructed on $(v_1, v_2)$
by Henneberg-I steps and only $deg(v_1)$ is 2. The recursive pattern is: for Case 2, 
$G\setminus \{v_1,v_2\}$ is also  1-path triangle-free graph which has low sampling complexity
on base non-edge $f=(v_3,v_4)$; for Case 3, $G\setminus \{v_1\}$ is also  
1-path triangle-free Henneberg-I graph which has simple sampling complexity
on base non-edge $f=(v_3,v_4)$. The quadrilateral structure for Case 2 is 
clear(see Figure~\ref{F:lemma-equivalence} (b)). 
For Case 3, we can see the quadrilateral structure by analyzing  $G\setminus \{v_1\}$
(see Figure~\ref{F:lemma-equivalence} (b)),
which already has one vertex $v_2$ which is constructed
on base non-edge $(v_3,v_4)$ and we know $deg(v_2)$ is not 2.
Without loss of generality we use $v_5$ to denote   
the other vertex
constructed on $(v_3, v_4)$ by a Henneberg-I step. In $G\setminus \{v_1\}$, 
if both $deg(v_3)$ and $deg(v_4)$ are 2(corresponding to Case 2), then we have
two quadrilaterals $(v_1,v_2,v_3,v_4)$ and $(v_2,v_3,v_4,v_5)$ which share
two edges $(v_2,v_3)$ and $(v_2,v_4)$(refer to Figure~\ref{F:triangle-free-basecases} (c)). 
In $G\setminus \{v_1\}$,
if only $deg(v_4)$ is 2 (corresponding to Case 1), then we
also have two quadrilaterals $(v_1,v_2,v_3,v_4)$ and $(v_2,v_3,v_4,v_5)$ which also share
two edges $(v_2,v_3)$ and $(v_2,v_4)$ (refer to Figure~\ref{F:triangle-free-basecases} (g)).
Since we can recursively repeat this analysis, 
if $G$ has low sampling complexity on $f$, $\Phi^2_f(G,\delta)$
can be computed by an $O(|V|)$ sequence of quadrilateral diagonal interval mappings.
\end{proof}

\subsubsection{Tightness of the forbidden minor characterization}

\begin{figure}[h]
\psfrag{1}{$v_1$}
\psfrag{2}{$v_2$}
\psfrag{3}{$v_3$}
\psfrag{4}{$v_4$}
\psfrag{5}{$v_5$}
\psfrag{6}{$v_6$}
\psfrag{7}{$v_7$}
\psfrag{8}{$v_8$}
\psfrag{9}{$v_9$}
\psfrag{10}{$v_{10}$}
\psfrag{11}{$v_{11}$}
\psfrag{12}{$v_{12}$}
\psfrag{13}{$v_{13}$}
\psfrag{14}{$v_{14}$}
\psfrag{g1}{$G_1$}
\psfrag{g2}{$G_2$}
\begin{center}
\includegraphics[width=10cm]{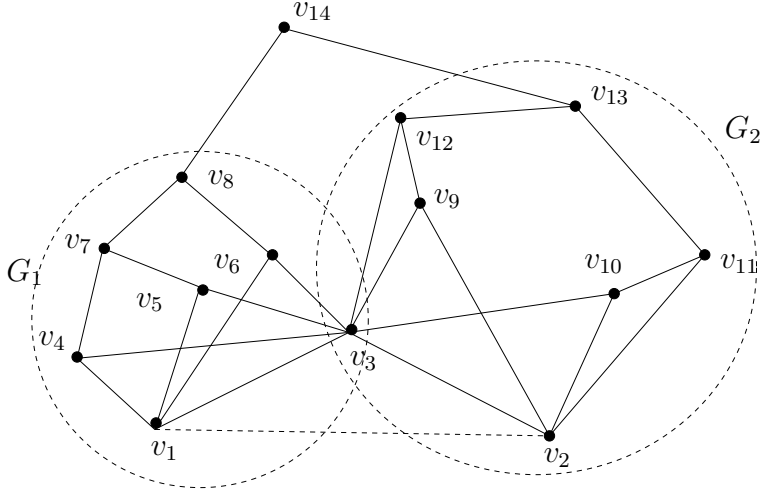}
\end{center}
\caption{
For Observation~\ref{obs:TriangleFreeCounter}.
A Simple 1-dof Henneberg-I graph $G$ on base non-edge
$(v_1,v_2)$ which is not triangle-free but 
has a single Henneberg-I construction
path for $v_{14}$ on base non-edge $(v_1,v_2)$; 
$G$ has configuration space of low sampling complexity on
$(v_1,v_2)$; but
$G_1$ has a
$K_{3,3}$ minor and $G_2$ has a $C_{3} \times C_{2}$ minor.
}
\label{F:TriangleFreeCounter}
\end{figure}

\begin{figure}[h]
\psfrag{1}{$v_1$}
\psfrag{2}{$v_2$}
\psfrag{3}{$v_3$}
\psfrag{4}{$v_4$}
\psfrag{5}{$v_5$}
\psfrag{6}{$v_6$}
\psfrag{7}{$v_7$}
\psfrag{8}{$v_8$}
\psfrag{9}{$v_9$}
\psfrag{10}{$v_{10}$}
\psfrag{11}{$v_{11}$}
\psfrag{12}{$v_{12}$}
\psfrag{13}{$v_{13}$}
\psfrag{14}{$v_{14}$}
\psfrag{g1}{$G_1$}
\psfrag{g2}{$G_2$}
\begin{center}
\includegraphics[width=10cm]{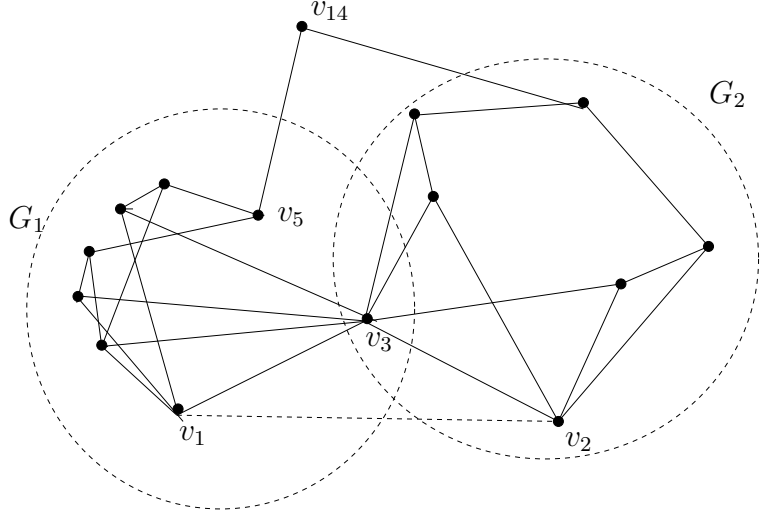}
\end{center}
\caption{
For Observation~\ref{obs:GeneralTriCounter}.
A Simple 1-dof Henneberg-I graph on base non-edge
$(v_1,v_2)$ that has one
Henneberg-I construction path on base non-edge $(v_1,v_2)$;
it has a configuration space of low sampling complexity on $(v_1,v_2)$ but
it has a $K_{5}$ minor shown in
the left circled subgraph; in general, it can have a arbitrary clique as
a minor.
}
\label{F:TriangleFreeCounter1}
\end{figure}

\begin{figure}[h]
\psfrag{1}{$v_1$}
\psfrag{2}{$v_2$}
\psfrag{3}{$v_3$}
\psfrag{4}{$v_4$}
\psfrag{5}{$v_5$}
\psfrag{6}{$v_6$}
\psfrag{7}{$v_7$}
\psfrag{8}{$v_8$}
\psfrag{9}{$v_9$}
\psfrag{10}{$v_{10}$}
\psfrag{11}{$v_{11}$}
\psfrag{12}{$v_{12}$}
\psfrag{13}{$v_{13}$}
\psfrag{14}{$v_{14}$}
\begin{center}
\includegraphics[width=7.5cm]{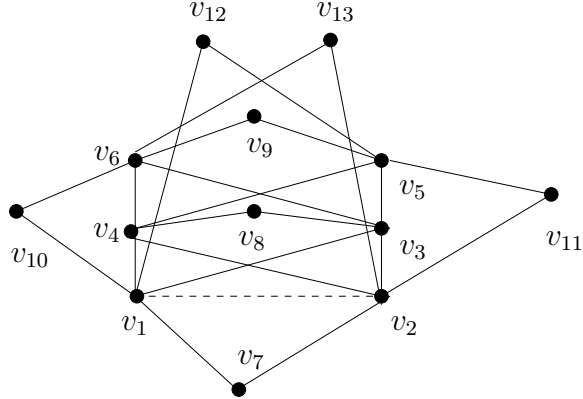}
\end{center}
\caption{
For Observation~\ref{obs:onepathcounter}.
A Simple 1-dof Henneberg-I graph on base non-edge
$(v_1,v_2)$ that has more than one 
Henneberg-I construction paths on base non-edge $(v_1,v_2)$;
it has a configuration space of low sampling complexity on $(v_1,v_2)$;
but it has both $K_{3,3}$ and $C_{3} \times C_{2}$ minors. Aside:
$(v_3,v_{4})$,
$(v_{5},v_{6})$, $(v_1,v_{5})$ and $(v_{2},v_{6})$
are also base non-edges and all of them yield configuration spaces of low sampling complexity.
}
\label{F:OnePathCounter}
\end{figure}

The following observations 
show that the characterization of Theorem \ref{the:TriangleFreeCase} is tight
by illustrating obstacles to obtaining a forbidden-minor
characterization after removing either of 
the restrictions of 
triangle-free(
Figures~\ref{F:TriangleFreeCounter} and \ref{F:TriangleFreeCounter1})
and 1-path(Figure~\ref{F:OnePathCounter})  used in the theorem.

\begin{observation}
\label{obs:TriangleFreeCounter}
There exists a 1-path Simple 1-dof Henneberg-I graph $G$ with base non-edge $f$
such that $G$ has low sampling complexity on $f$ but $G$ has both $K_{3,3}$
and $C_{3} \times C_{2}$ minor.
\end{observation}

\noindent
\begin{proof}
We give such a graph $G$ in Figure~\ref{F:TriangleFreeCounter}. We can verify 
that $G$ is a 1-path Simple 1-dof Henneberg-I graph with base non-edge $f=(v_1,v_2)$.
Among all the extreme graphs corresponding to the Henneberg-I steps, only 
$G \cup (v_1,v_2)$ and $G \cup (v_8,v_{13})$ are well-constrained. Since
$G \cup (v_1,v_2)$ is a Henneberg-I graph with base edge $(v_1,v_2)$, it follows that 
$G \cup (v_1,v_2)$ is Triangle-decomposable. Now that the subgraph induced
by $\{v_1, v_3, v_4, v_5, v_6, v_7, v_8\}$ is a 1-path Henneberg-I graph
with base edge $(v_1,v_3)$, which we denote by $G_1$. The subgraph induced
by $\{v_2, v_3, v_9, v_{10}, v_{11}, v_{12}, v_{13}\}$ is a 1-path Henneberg-I graph
with base edge $(v_1,v_3)$, which we denote by $G_2$. So the wellconstrained extreme graph
$G \cup (v_8,v_{13})$ has a Triangle-decomposition, 
$G_1$, $G_2$ and  $\triangle(v_8,v_{13},v_{14})$. Because all the wellconstrained extreme 
graphs ($G \cup (v_1,v_2)$ and $G \cup (v_8,v_{13})$ ) of 
$G$ are Triangle-decomposable, $G$ 
has low sampling complexity on $f$ by the definiton of low sampling complexity.
It is clear that $G_1$ has a $K_{3,3}$ minor and $G_2$ has a $C_3 \times C_2$ minor,
so the example we have constructed satisfies all the requirements.
\end{proof}

By minor modification of Figure~\ref{F:TriangleFreeCounter}, 
we have the following stronger observation.

\begin{observation}
\label{obs:GeneralTriCounter}
For any graph $G_s$, there exists a 1-path Simple 1-dof Henneberg-I graph $G$ with base non-edge $f$
such that $G$ has low sampling complexity on $f$ and $G_s$ is a minor of $G$.
\end{observation}

\noindent
\begin{proof}
We only need to prove the case that $G_s$ is an arbitrary clique $K_m$. 
To do that,
we change the subgraph $G_1$ in Figure~\ref{F:TriangleFreeCounter} 
such that $K_m$ is a minor of $G_1$. If we can do this, the proof follows since
by using the same argument as proof for Observation~\ref{obs:TriangleFreeCounter},
we additionally have: both $G_1$ 
and $G_2$ are Henneberg-I graphs with base edge $(v_1,v_3)$ and $(v_2,v_3)$ 
respectively; there are only two extreme graphs which are both wellconstrained and
Triangle-decomposable so $G$ has low sampling complexity on $(v_1,v_2)$.

We prove by induction that we can construct a 1-path Henneberg-I graph 
$G_1$ with base edge $(v_1,v_3)$ such that $G_1$ 
can be reduced to $K_m$ by edge contractions. The base cases ($m = 1,2,3$) have 
been shown in Figure~\ref{F:TriangleFreeCounter}. As the induction
hypothesis, we assume that we can construct a 1-path Henneberg-I graph 
$G_1^m$ with base edge $(v_1,v_3)$ such that $G_1^m$ 
can be reduced to $K_m$ by edge contractions. Now we prove the induction step for $K_{m+1}$.
We start from $G_1^m$ to construct $G_1^{m+1}$. We pick $m$ vertices $u_1, \cdots, u_m$
from $G_1^m$  containing the last constructed vertex of $G_1^m$ and additionally such that
they map to distinct vertices 
in the contracted graph $K_m$. We add a
vertex $w_1$ by a Henneberg-I step with $u_1$ and $u_2$ as base vertices. 
Then we add a vertex $w_2$ by a Henneberg-I step with $w_1$ and $u_2$ as base vertices
and so on. Finally we add $w_{m-1}$ by Henneberg-I step with $w_{m-2}$ and $u_m$
as base vertices to get $G_1^{m+1}$ (Please refer to Figure~\ref{F:TriangleFreeCounter1} for a $K_{5}$
example). Clearly, $G_1^{m+1}$ is
a Henneberg-I graph with  base edge $(v_1,v_3)$.
Then by contracting all the edges that 
have at least one vertex other than $u_1$, $\cdots$, $u_m$ and $w_{m-1}$,
we get a $K_{m+1}$. Thus, we have proved that $G_1^{m+1}$ is a Henneberg-I graph and can 
be contracted to $K_{m+1}$. 
\end{proof}

\begin{observation}
\label{obs:onepathcounter}
There exists a triangle-free Simple 1-dof Henneberg-I graph
with base non-edge $f=(v_1,v_2)$ such that $G$ has low sampling 
complexity on $f$ and $G$ has both $K_{3,3}$ and/or $C_{3} \times C_{2}$ minor.
\end{observation}

\noindent
\begin{proof}
We give such a graph $G$ in Figure~\ref{F:OnePathCounter}. 
The Simple 1-dof Henneberg-I graph is constructed with base non-edge
$(v_1,v_2)$ and it is not a 1-path. We can verify that all the 
extreme graphs are in fact Henneberg-I graphs so they are 
Triangle-decomposable. This shows that $G$ has low sampling complexity
on $f$. If we contract all the edges that have at least one vertex
other than $v_1$, $v_2$, $v_3$, $v_4$, $v_5$ and $v_6$, we can get
a clique $K_6$, so $G$ has both $K_{3,3}$ and $C_{3} \times C_{2}$ minors.
$G$ has all the required properties of the observation.
\end{proof}

\subsubsection{Graph characterization for 1-path Simple 1-dof Henneberg-I graph}

Despite the obstacles to obtaining a forbidden minor characterization when the ``triangle-free''
restriction is removed, 
the following theorem gives a characterization of
 1-path 1-dof  Simple Henneberg-I graphs $G$ 
that have low sampling complexity.


\begin{figure}[h]
\psfrag{1}{$v_1$}
\psfrag{2}{$v_2$}
\psfrag{3}{$v_3$}
\psfrag{4}{$v_4$}
\psfrag{5}{$v_5$}
\psfrag{6}{$v_6$}
\psfrag{7}{$v_7$}
\psfrag{8}{$v_8$}
\psfrag{9}{$v_9$}
\psfrag{10}{$v_{10}$}
\psfrag{11}{$v_{11}$}
\psfrag{12}{$v_{12}$}
\psfrag{13}{$v_{13}$}
\psfrag{14}{$v_{14}$}
\psfrag{15}{$v_{15}$}
\psfrag{16}{$v_{16}$}
\psfrag{17}{$v_{17}$}
\psfrag{18}{$v_{18}$}
\psfrag{19}{$v_{19}$}
\psfrag{20}{$v_{20}$}
\psfrag{21}{$v_{21}$}
\psfrag{22}{$v_{22}$}
\psfrag{23}{$v_{23}$}
\psfrag{24}{$v_{24}$}
\psfrag{25}{$v_{25}$}
\psfrag{26}{$v_{26}$}
\psfrag{27}{$v_{27}$}
\begin{center}
\includegraphics[width=7.5cm]{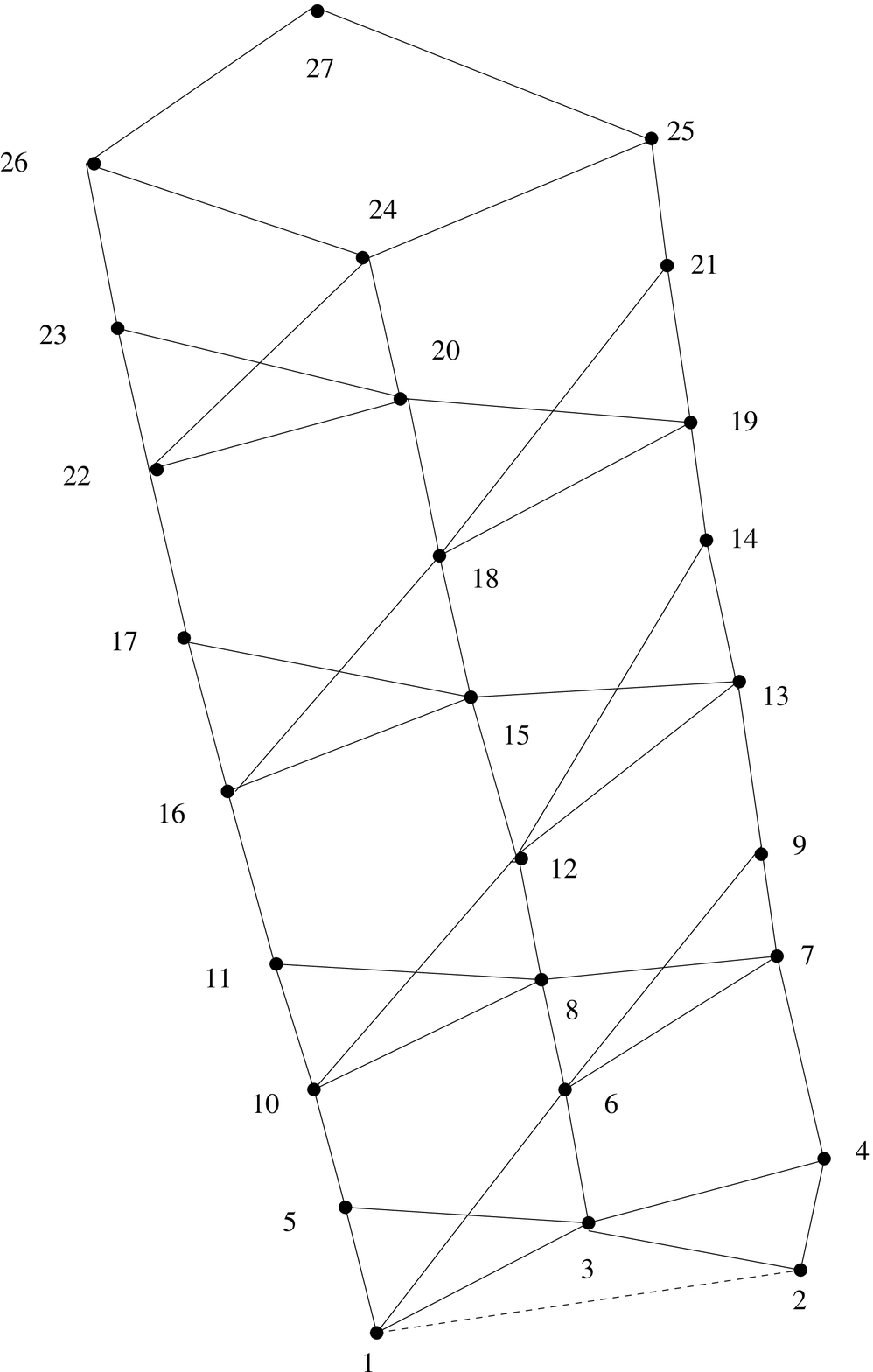}
\end{center}
\caption{
A 1-path 1-dof Henneberg-I graph that has low sampling complexity on
base non-edge $(v_1,v_2)$;
exactly 1 vertex namely $v_3$ is constructed on $v_1$
and $v_2$. See proof of Theorem~\ref{the:chain} and Observation~\ref{obs:one-path-not-direct}.
}
\label{F:GoodRCC}
\end{figure}

\begin{theorem} \label{the:chain}
Given a 1-path Simple 1-dof Henneberg-I graph $G$ with specified
base non-edge $(v_1,v_2)$, if $G$ has low sampling complexity on $v_1, v_2$
then
the number of vertices directly constructed using $v_1$ and $v_2$
as base vertices is 1 or 2.
If it is 2,
$G$ has 
low sampling complexity
 on $(v_1,v_2)$
if and only if the following hold:
(1)
either $v_1$
or $v_2$ has a degree of 2; (2) if $v_3$ and $v_4$ are
the only vertices directly constructed
on $v_1$ and $v_2$ and the degrees of both $v_1$ and $v_2$ are 2,
1-path Simple 1-dof Henneberg-I graph $G \setminus \{v_1,v_{2}\}$ must
have low sampling complexity
on base non-edge
$(v_3,v_4)$;
(3) if $v_3$ and $v_4$ are
the only vertices directly constructed
on $v_1$ and $v_2$ and only one of $v_1$ and $v_2$ has degree of 2, without loss
of generality say $v_2$, then
$G \setminus \{v_2\}$ has to
be a Simple 1-dof 1-Path Henneberg-I graph that has 
low sampling complexity
on base non-edge $(v_3,v_4)$.
\end{theorem}

\noindent
\begin{proof}
For (1), 
assume there are $m$ vertices in $G$ that are directly constructed 
with $v_1$ and $v_2$ as base vertices. By Lemma~\ref{lem:4lemmas} (1.a), 
$m < 3$. When $m = 2$, assume  $v_3$ and $v_4$ are constructed
with $v_1$ and $v_2$ as base vertices. Since $v_1$ and $v_2$ are adjacent to 
both $v_3$ and $v_4$, so both $deg(v_1)$ and $deg(v_2)$ are at least 2.  
By Lemma~\ref{lem:4lemmas} 1.b, $deg(v_1)$ and $deg(v_2)$ cannot be both greater
than 2, so either $v_1$ or $v_2$ has degree of two. 

By Lemma~\ref{lem:equivalence} (1.a) and (1.b) we have (2) and
by Lemma~\ref{lem:equivalence} (2.a) and (2.b) we have (3).
\end{proof}

\medskip
{\bf Remark:} Theorem~\ref{the:chain} leaves the case where the number of vertices 
directly constructed using $v_1$ and $v_2$ as base vertices is exactly 1.
Such graphs of low sampling complexity are captured in Figure~\ref{F:GoodRCC}. 
The graph characterization for this type of graphs is complicated 
and appears in \cite{bib:Gao08}.



\medskip
The above theorem characterizes 
the Simple, 1-dof, 1-path Henneberg-I graphs that have 
Triangle-decomposable extreme graphs and low sampling
complexity. It is natural to ask if the configuration
space description for these graphs can also be obtained
directly as in Observation \ref{obs:backward}, without actually
realizing the extreme graphs. 
The next observation gives a negative answer.

\begin{observation}
\label{obs:one-path-not-direct}
Figure \ref{F:GoodRCC} shows
an example of a Simple, 1-dof, 1-path Henneberg-I graph with
low sampling complexity, for which the interval endpoints in its configuration
space cannot directly be obtained by the method of quadrilateral diagonal interval mapping
(in Observation~\ref{obs:backward}).
\end{observation}

\noindent
\begin{proof}
In Figure~\ref{F:GoodRCC}, the graph is a 1-path 1-dof Henneberg-I graph that
has low sampling complexity on base non-edge $(v_1,v_2)$. However, we cannot 
find a sequence of quadrilaterals such that we can use a quadrilateral diagonal interval mapping. 
For example, if we start from $\triangle (v_{25},v_{26},v_{27})$ and get an
interval for $\delta^*(v_{25},v_{26})$, then we have no further quadrilaterals to proceed.
\end{proof}

\subsection{Characterizing parameter choices: all base edges yield equally efficient configuration spaces}

We show an interesting quantifier exchange theorem
for Henneberg-I graphs. Besides providing
a characterization of all possible parameters that yield efficient
configuration spaces, 
the theorem illustrates the robustness of 
our definition of low sampling complexity.

\begin{theorem}
\label{the:quantifierExchange}
For a Henneberg-I graph $H$, consider each possible base edge $f$ that
gives a Henneberg construction for $H$.
Let $H_f$ be the Simple 1-dof Henneberg-I graph with base non-edge $f$.
Then either $\Phi_f^2(H_f,\delta)$ has low sampling complexity for all
base edges $f$ of $H$ or 
for none of them.
See Figure \ref{F:OnePathCounter}.
\end{theorem}

\noindent
\begin{proof}
We prove by contradiction. Suppose the proposition is not true, then we can find
ив set of Simple 1-dof Henneberg-I graphs such that for each of them we can find two
base non-edges such that it has  low sampling complexity on one base non-edge
while does not have sampling complexity on the other. We pick a graph 
$G$ among this set such that the number of the vertices of $G$ is the minimum. 
If there is any tie, we break the tie arbitrarily. Without loss of generality, we assume that 
$G$ has two base non-edges $f_1 = (v_1, v_2)$ and $f_2 = (v_3, v_4)$ and
$G$ has low sampling complexity on $f_1$ but does not have low sampling complexity
on $f_2$. Clearly $f_1$ and $f_2$ cannot be the same and $G$ has at least
3 vertices.

The proof rests on several claims on the above $G$, $f_1$ and $f_2$ which exclude the
possibility of minimality.

\begin{claim} 
\label{claim1}
For any vertex of $G$ other than $v_1$, $v_2$, $v_3$ and $v_4$
cannot have degree 2. 
\end{claim}

\noindent
\begin{proof}
To the contrary, suppose a vertex $v_{n}$ has degree
2. We used $G\p$ to denote the graph we get by removing $v_n$
from $G$. Without loss of generality, suppose $v_{n}$ is constructed with
$u_n$ and $w_n$ as base vertices. If $G$ has a wellconstrained subgraph which
includes both $u_n$ and $w_n$, by Fact~\ref{fact:wellcondition} the subgraph 
is a Henneberg-I graph and also a 2-sum component of $G$. 
So $G\p$ will be also a Henneberg-I 
graph with two base non-edges $f_1$ and $f_2$ and just like $G$, 
$G\p$ has low sampling complexity on $f_1$ but does not have low sampling complexity
on $f_2$. Now we consider the case that $G$ does not have a wellconstrained subgraph which
include both $u_n$ and $w_n$. In this case, since $G$ has low sampling complexity on $f_1$,
it follows that $G \cup (u_n,w_n)$ is Triangle-decomposable. 
Compare the extreme graphs associated with $G$ and $(v_3, v_4)$ and
the extreme graphs associated with $G\p$ and $(v_3, v_4)$, the former
has one more extreme graph $G \cup (v_1, v_2)$. $G$ does not have
low sampling complexity on $f_2 = (v_3, v_4)$, so one extreme graph
associated with $(G, f_2)$ is not Triangle-Decomposable. 
Now $G\cup (v_i, v_j)$ is Triangle-Decomposable, so one extreme graph
associated with $(G\p, f_2)$
must not be Triangle-Decomposable, thus, $G\p$ must have one extreme
graph which is not Triangle-Decomposable such that $G\p$ does not
have low sampling complexity on $f_2$.
In both cases, $G\p$ is a Henneberg-I 
graph with base non-edges $f_1$ and $f_2$ and just like $G$, 
$G\p$ has low sampling complexity on $f_1$ but does not have low sampling complexity
on $f_2$. This violates the minimality of $G$, so our assumption is incorrect and
no vertex of $G$ other than $v_1$, $v_2$, $v_3$ and $v_4$
can have degree 2.
\end{proof}

\begin{claim}
\label{claim2}
At least one of $deg(v_1)$ and $deg(v_{2})$ is 2;
similarly, at least one of $deg(v_{3})$ and $deg(v_{4})$ is 2. 
\end{claim}

\noindent
\begin{proof}
 Since $G \cup f_1$
is a Henneberg-I graph with base non-edge $(v_1, v_2)$(and $G$ has
at least 3 vertices), so there is at least one vertex other than $v_1$ and $v_2$ 
which has degree of 2. From (1), any vertex other than  $v_1$, $v_2$, $v_3$ and $v_4$
cannot have degree 2, so either $deg(v_3)$ or $deg(v_{4})$ is 2 since
$v_3$ and $v_4$ are the only vertices other than $v_1$ and $v_2$ which can have degree 2.
Similarly, at least one of $deg(v_3)$ or $deg_{v_4}$ is 2. Without loss of generality we suppose that
$deg(v_1)$ and $deg(v_3)$ are 2.
\end{proof}

\begin{claim}
\label{claim3}
There is 
only one vertex constructed with $v_1$ and $v_2$ as base vertices.
\end{claim}

\noindent
\begin{proof}
In Claim~\ref{claim2}, we proved that at least one of $deg(v_1)$ and $deg(v_{2})$ is two
we assume that $deg(v_1)$ is 2
so the number of  vertices constructed with $v_1$ and $v_2$
as base vertices is at most two. We can show by contradiction that this number 
is not exactly 2. Suppose 
$v_{5}$ and $v_{6}$ are the two vertices constructed with $v_1$ and $v_2$ as
base vertices. 
By Lemma~\ref{lem:equivalence}(2.a), $(v_5, v_6)$ is also a base
non-edge for $G$. $G$ has low sampling complexity on $(v_1, v_2)$, so by
Lemma~\ref{lem:equivalence}(1.b or 2.b), $G$ also has low sampling complexity on $(v_5, v_6)$.
If $v_1$ is different from both $v_3$ and $v_4$, 
we have: $G$ does not have low sampling complexity
on $(v_3, v_4)$,  $G$ has low sampling complexity on $(v_5, v_6)$,
$v_1$ has degree 2, and $v_1$ is different from 
$v_3$, $v_4$, $v_5$ and $v_6$. This contradicts to Cliam~\ref{claim1}, 
so we only need to consider the case that $v_1$ is the same as $v_3$ or $v_4$. 
Without loss of generality, we assume that $v_1$ is the same as $v_3$. 
$G$ is a 1-dof Henneberg-I graph with base non-edge $(v_3, v_4)$ and $G$ 
has at least 3 vertices, so at least one vertex of $G$ other than 
$v_3$ and $v_4$ has degree 2. By Cliam~\ref{claim1}, only $v_1$, $v_2$, $v_3$
and $v_4$ can have degree 2, so $deg(v_2)$ has to be 2. Now,
$G$ has low sampling complexity 
on $(v_5, v_6)$, $G$ does not have low sampling complexity 
on $(v_3, v_4)$, vertex $v_2$ has degree of 2 and $v_2$ is
different from $v_3$, $v_4$, $v_5$ and $v_6$. This again contradicts
to Cliam~\ref{claim1} thus proves the claim.
\end{proof}

\medskip
{\bf [Theorem~\ref{the:quantifierExchange} Continued]}
  By Claim~\ref{claim3} there is only one vertex constructed with
$v_1$ and $v_2$ as base vertices, without loss of generality, suppose $v_9$ is such 
a vertex. Consider the Henneberg-I step that immediately 
follows $v_9 \triangleleft (v_1, v_2)$. Since $v_9$
is the vertex constructed with
$v_1$ and $v_2$ as base vertices, the base vertices for the next
Henneberg-I step are either $v_1$ and $v_9$ or $v_2$ and $v_9$.
Since we have labeled $v_1$ as the vertex which has degree of 2,
we have to differentiate these two cases. In Claim~\ref{claim5} we discuss the case
in which the next Henneberg-I step is 
$v_{10}\triangleleft (v_1,v_9)$ and in Claim~\ref{claim6} we discuss the case
in which the next Henneberg-I step is 
$v_{10}\triangleleft (v_2,v_9)$.

\begin{claim}
\label{claim5}
If $v_9$ is the only vertex constructed with $v_1$ and $v_2$ as base vertices and
$deg(v_1)$ is 2, then no vertex can be constructed with $v_1$ and $v_9$ as 
base vertices.
\end{claim}

\begin{figure}[h]
\psfrag{1}{$v_1$}
\psfrag{2}{$v_2$}
\psfrag{3}{$v_3$}
\psfrag{4}{$v_4$}
\psfrag{5}{$v_5$}
\psfrag{6}{$v_6$}
\psfrag{7}{$v_7$}
\psfrag{8}{$v_8$}
\psfrag{9}{$v_9$}
\psfrag{10}{$v_{10}$}
\psfrag{11}{$v_{11}$}
\psfrag{12}{$v_{12}$}
\psfrag{13}{$v_{13}$}
\psfrag{14}{$v_{14}$}
\psfrag{1,4}{$v_1$($v_4$)}
\begin{center}
\includegraphics[width=12cm]{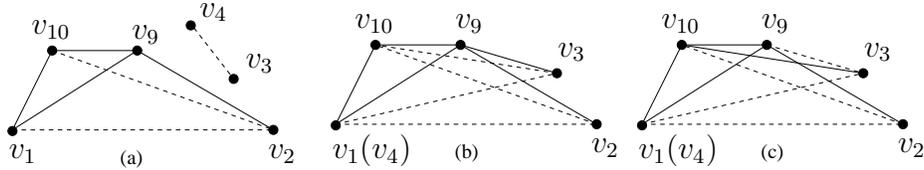}
\end{center}
\caption{
Proofs of Claim~\ref{claim5} and in Claim~\ref{claim6}.
Simple 1-dof Henneberg-I graph $G$ has low sampling complexity on base non-edge $(v_1,v_2)$
while $G$ does not have low sampling complexity on base non-edge $(v_3,v_4)$;
vertex $v_9$ is the only vertex directly constructed on $v_1$ and $v_2$;
triangle $\triangle (v_9, v_{10}, v_1)$ corresponds to the second Henneberg-I construction
from $(v_1,v_2)$; in (a), $(v_1,v_2)$ and
$(v_3,v_4)$ do not share any vertex; in (b) and (c), $(v_1,v_2)$ and
$(v_3,v_4)$ share a vertex.
}
\label{F:aaa}
\end{figure}

\noindent
\begin{proof}  
Refer to Figure~\ref{F:aaa}, $(v_2, v_{10})$ is also 
a base non-edge for Simple 1-dof Henneberg-I graph $G$. 
Further $G$ has low sampling complexity on $(v_2, v_{10})$
since  $G$ has low sampling complexity on $(v_1, v_2)$. 
This is a result which is similar to Lemma~\ref{lem:equivalence} (1.b) and
can be proved by comparing all the possible extreme graphs. 
For any extreme graph
corresponding to $(v_{2},v_{10})$, there is an extreme graph
corresponding to $(v_{1},v_{2})$ which has one extra Henneberg-I step
$v_1 \triangleleft (v_9, v_{10})$. 
$G$ has low sampling complexity on $(v_1, v_2)$, so all the extreme graphs
corresponding to  $(v_{1},v_{2})$ are triangle decomposable. 
Thus,
all the extreme graphs
corresponding to  $(v_2,v_{10})$ are also triangle decomposable
since removing vertices from a triangle decomposable graph by inverse Henneberg-I steps
keeps the graph still triangle decomposable. So, $G$ has low sampling 
complexity on $(v_{2},v_{10})$.

Note that $G$ has does not have low sampling complexity on $(v_3,v_4)$ and
$deg(v_1)$ is 2. By Claim~\ref{claim1},  $v_1$ cannot be different from both $v_3$ and $v_4$ 
(Figure~\ref{F:aaa}(a)). So we only need to consider the case $v_1$ 
is coincident with $v_3$ or $v_4$. Although we labeled $v_3$ as the
vertex with degree of 2  but we do not use this property here, so we suppose $v_1$ 
is coincident with $v_4$. 

Since $(v_3, v_4)$ is a base non-edge for $G$ and
$v_4$ (just like $v_1$) is only adjacent to $v_9$ and $v_{10}$, so $v_3$
must be adjacent to either $v_9$ (Figure~\ref{F:aaa}(b))  or
$v_{10}$ (Figure~\ref{F:aaa}(c)) in order to guarantee the Henneberg-I step
with $v_3$ and $v_4$ as base vertices is possible.
For Figure~\ref{F:aaa}(b), $(v_3,v_{10})$ is a base non-edge
for $G$ since  $(v_3,v_{4})$ is a base non-edge
for $G$. If we compare the extreme graphs corresponding to  $(v_3,v_{10})$
with  the extreme graphs corresponding to  $(v_3,v_4)$, the only difference
is: the extreme graph in the  latter case has one more Henneberg-I step
$v_4 \triangleleft (v_9,v_{10})$. Note that removing/adding a Henneberg-I step
to a graph does not change
triangle decomposability. Now $G$ does not have low sampling complexity on
 $(v_3,v_4)$, so $G$ does not have low sampling complexity on
 $(v_3,v_{10})$. Now we have:  $G$ does not have low sampling complexity on
 $(v_3,v_{10})$,  $G$ has low sampling complexity on
 $(v_1,v_2)$, $v_1$ has degree of 2 and $v_1$ is different from
 $v_2$, $v_3$, $v_9$ and $v_{10}$. This contradicts to the result in Claim~\ref{claim1},
 so the case shown in Figure~\ref{F:aaa}(b) is impossible. 
 We can use the same argument for the case shown in
 Figure~\ref{F:aaa}(c) and get: $G$ does not have low sampling complexity on
 $(v_3,v_{9})$,  $G$ has low sampling complexity on
 $(v_1,v_2)$, $v_1$ has degree of 2 and $v_1$ is different from
 $v_2$, $v_3$, $v_9$ and $v_{10}$. This again contradicts to Claim~\ref{claim1},
 so the case shown in Figure~\ref{F:aaa}(c) is also impossible. 
 Now we have shown that we cannot have a Henneberg-I step $v_{10}\triangleleft (v_1,v_9)$. 
\end{proof}



\begin{claim}
\label{claim6}
If $v_9$ is the only vertex constructed with $v_1$ and $v_2$ as base vertices and
$deg(v_1)$ is 2, then no vertex can be constructed with $v_2$ and $v_9$ as 
base vertices either.
\end{claim}

\noindent
\begin{proof}
Otherwise, let $v_{10}$ 
be constructed with $v_2$ and $v_9$ as base vertices(see Figure~\ref{F:bbb}
and Figure~\ref{F:ccc}). 

Let $v_{12}$ denote the other vertex that $v_1$ is adjacent 
(we know that $v_1$ is adjacent to $v_9$). 
Observe that $(v_1, v_2)$ is a base non-edge for $G$, so $v_1$ must be one
of the two base vertices for $v_{12}$'s construction.
Denote the other vertex by $v_{11}$. Clearly, before $v_{12}$
is constructed, we must have constructed a 1-path Henneberg-I graph with 
$(v_2, v_9)$ as base edge and $v_{11}$ as the last vertex. We denote
this 1-path Henneberg-I graph by $G_1$ and mark it by a dashed circle
in Figure~\ref{F:bbb}.

\begin{figure}[h]

\begin{center}
\psfrag{1}{$v_1$}
\psfrag{2}{$v_2$}
\psfrag{3}{$v_3$}
\psfrag{4}{$v_4$}
\psfrag{5}{$v_5$}
\psfrag{6}{$v_6$}
\psfrag{7}{$v_7$}
\psfrag{8}{$v_8$}
\psfrag{9}{$v_9$}
\psfrag{10}{$v_{10}$}
\psfrag{11}{$v_{11}$}
\psfrag{12}{$v_{12}$}
\psfrag{13}{$v_{13}$}
\psfrag{14}{$v_{14}$}
\psfrag{1,3}{$v_1$($v_3$)}
\psfrag{g1}{$G_1$}
\includegraphics[width=6cm]{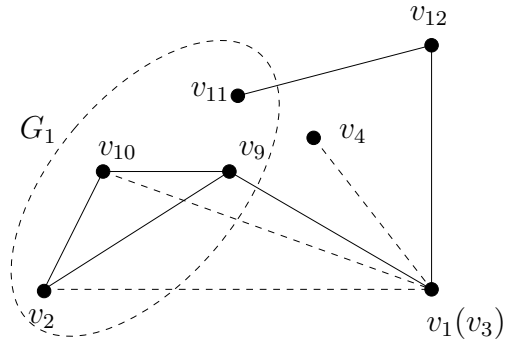}
\end{center}
\caption{Proof of Claim~\ref{claim6}: $v_1$ is coincident with $v_3$.}
\label{F:bbb}
\end{figure}

We can show that $v_1$ has to be different from  $v_3$ and $v_4$. Since $G$ is simple 1-dof Henneberg-I
graph with base non-edge $(v_3,v_4)$, at least one vertex other than
$v_3$ and $v_4$ should have degree  2.  By Claim~\ref{claim1}, $v_1$ and $v_2$ are the only
possible vertices with degree of 2. If $v_1$ is $v_3$ or $v_4$, 
$deg(v_2)$ is 2. But $deg(v_2)$ cannot be 2 by Claim~\ref{claim5},
so we have proved that $v_1$ has to be different from  $v_3$ and $v_4$
and $deg(v_2)$ is not 2(see Figure~\ref{F:bbb}).

For the remaining cases, we will use Claim~\ref{claim1} to target
the impossbility which is stated in the laim we want to prove. To 
do that, we change the edges of $G$ to get a new graph $G$ such that: 
$G\p$ has low sampling complexity on base non-edge $f_3$; 
$G\p$ does not have low sampling complexity on base non-edge $(v_3, v_4)$;
$v_1$ is different from $v_3$, $v_4$ and the two vertices of $f_3$.   

If we consider the Henneberg-I sequence
starting from $(v_1, v_2)$, $G_1$ is a Henneberg-I graph with
$(v_2, v_9)$ as base vertice and the last vertex is $v_{11}$.
This means that any vertex in $G_1$ other than $v_2$, $v_9$ and
$v_{11}$ do not have degree 2. Consider how we
can construct $G_1$ in the Henneberg-I sequence starting from
$(v_3, v_4)$. Recall each Henneberg-I step involves 
1 vertex and 2 edges. By using the same dof counting method that used 
for Fact~\ref{fact:wellcondition}, there must be
an edge between the first two vertices in $G_1$, without loss of generality we assume that
the first vertex is $v_{13}$ and the second is $v_{14}$. So, $G_1$ 
has to be a Henneberg-I graph(may not be 1-path) with base edge $(v_{13}, v_{14})$.

\begin{figure}[h]
\psfrag{1}{$v_1$}
\psfrag{2}{$v_2$}
\psfrag{3}{$v_3$}
\psfrag{4}{$v_4$}
\psfrag{5}{$v_5$}
\psfrag{6}{$v_6$}
\psfrag{7}{$v_7$}
\psfrag{8}{$v_8$}
\psfrag{9}{$v_9$}
\psfrag{10}{$v_{10}$}
\psfrag{11}{$v_{11}$}
\psfrag{12}{$v_{12}$}
\psfrag{13}{$v_{13}$}
\psfrag{14}{$v_{14}$}
\psfrag{15}{$v_{15}$}
\psfrag{g1}{$G_1$}
\begin{center}
\includegraphics[width=10cm]{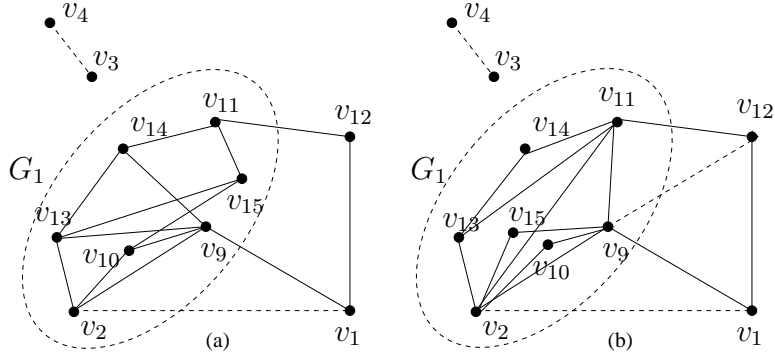}
\end{center}
\caption{For proof of Claim~\ref{claim6}: $v_1$ is different from $v_3$ and $v_4$.}
\label{F:ccc}
\end{figure}

Now we modify $G_1$ to get $G_1\p$ such that we get the $G\p$ that
we expect. We keep all the vertices in $G_1$ but remove all the edges.
Our objective is to add edges to get a new graph $G_1\p$ such that 
$G_1\p$ is a Henneberg-I graph with both $(v_2, v_9)$ and $(v_{13},v_{14})$
as base edges and $G_1\p$ contains edges $(v_2, v_{11})$ and $(v_9, v_{11})$.
To achieve this, we first add edges $(v_2, v_9)$, $(v_2, v_{11})$ and $(v_9, v_{11})$.
Then we consider adding edges for  $v_{13}$ and $v_{14}$: if both $v_{13}$ and $v_{14}$ are 
among $v_2$, $v_9$
or $v_{11}$, we do not add any edge;
if exactly one of $v_{13}$ and $v_{14}$ is one of $v_2$, $v_9$
or $v_{11}$, we add edges $(v_{13}, v_{14})$ and another edge between  $v_{11}$
and whichever of $v_{13}$ and $v_{14}$ is not one of  $v_2$, $v_9$
or $v_{11}$; if neither of $v_{13}$ and $v_{14}$
is one of $v_2$, $v_9$
or $v_{11}$, we add edges $(v_{13}, v_{2})$, $(v_{13}, v_{11})$ 
$(v_{14}, v_{13})$ and 
$(v_{14}, v_{11})$. Finally for each vertex $u$ in $G_1$ other than 
$v_2$, $v_9$, $v_{11}$, $v_{13}$ and $v_{14}$, we add 
one edge between $u$ and $v_2$ and another one
between $u$ and $v_9$. We use $G_1\p$ to denote this new subgraph that replaces $G_1$
and $G\p$ for the entire graph. By the manner in which we add edges, 
our objective is achieved:
 $G_1\p$ is Henneberg-I graph with both $(v_2, v_9)$ and $(v_{13}, v_{14})$
 as base edges and also contains edges  $(v_2, v_{11})$ and $(v_{9}, v_{11})$.

Now observe that both $(v_1, v_2)$ and $(v_3, v_4)$ are still
base non-edges for $G\p$. Further, $(v_9, v_{12})$ is also a
base non-edge for $G$. 
Now we consider whether $G\p$
has low sampling complexity on  $(v_9, v_{12})$  and  $(v_3, v_{4})$. 
To do that, we refer to Theorem 1 proved in \cite{bib:FudHo97proof}:
if a graph is Triangle-decomposable,
we can perform the cluster merging (inverse operation of 
triangle-decomposition) in any order(a Church-Rosser Property) but finally we get one cluster
which is the same as the whole graph. So for any given graph, if we replace one 
of its
Triangle-decomposable subgraphs by another triangle decomposable subgraph
while keeping the vertices unchanged, the graph preserves 
Triangle-decomposability. Here in our transform, 
both $G_1$ and $G_1\p$ are Henneberg-I graphs and thus both are 
Triangle-decomposable. Compare the extreme graphs corresponding to 
$G$ and $G\p$ for which base non-edge is chosen as $(v_1, v_2)$.

Observe that we are only interested in well-constrained extreme graphs. 
By Fact~\ref{fact:extremeWell}, an extreme graph corresponding to the
Henneberg-I step $v\triangleleft(u,w)$ is wellconstrained if and only if
$u$ and $w$ are not in any wellconstrained subgraph. Observe that the difference
between $G$ and $G\p$ is exactly the difference between 
$G_1$ and $G_1\p$. Both $G_1$ and $G_1\p$ are wellconstrained, so in the
comparison of extreme graphs we do not need to consider extreme graphs corresponding to
the Henneberg-I steps inside
$G_1$ and $G_1\p$. 

For all the other Henneberg-I steps outside $G_1$
and $G_1\p$, the difference between the extreme graphs for $G$ and $G\p$
is exactly the difference between $G_1$ and $G\p$. This proves 
$G\p$ has low sampling complexity on $(v_1,v_2)$ since 
$G$ has low sampling complexity on $(v_1, v_2)$.
Similarly, we can show that $G\p$ does not have 
 low sampling complexity on $(v_3,v_4)$ since 
$G$  does not have low sampling complexity on $(v_3, v_4)$.
Now verifying Figure~\ref{F:ccc} again, $(v_9, v_{12})$ is also
a base non-edge for $G\p$. By comparison of extreme graphs
as we did in Claim~\ref{claim5}, $G\p$ has low sampling complexity on $(v_1,v_2)$ since
$G\p$ has low sampling complexity on $(v_9, v_{12})$. 
This contradicts to Claim~\ref{claim1}, so we have proved
when $v_9$ is the only vertex constructed with $v_1$ and $v_2$ as base vertices and
$deg(v_1)$ is 2, then no vertex can be constructed with $v_2$ and $v_9$ as 
base vertices either.
\end{proof}

\medskip
{\bf [Theorem~\ref{the:quantifierExchange} Continued]}
Now we can put all the 5 claims together. 
We assume that $G$ has
low sampling complexity on base non-edge $(v_1, v_2)$ but
does not have low sampling complexity on base non-edge $(v_3, v_4)$. 
We also assume that the number of vertices in $G$ is minimum among
all the graphs with this property. In Claim~\ref{claim1} to Claim~\ref{claim6}, we discuss what
properties such a $G$ should have in order to keep the minimality
of the number of vertices. In Claim~\ref{claim1} we show that any vertex other than
$v_1$, $v_2$, $v_3$ and $v_4$ cannot have degree 2; in Claim~\ref{claim2}, 
we show at least one of $deg(v_1)$ and $deg(v_2)$
(resp. at least one of $deg(v_3)$ and $deg_{v_4}$) is 2 and without loss of generality
we assume that $deg(v_1)$ and $deg(v_{3})$ are 2;
in Claim~\ref{claim3}, we show that there is only vertex that is constructed
with $v_1$ and $v_2$ as base vertices and we denote the vertex
by $v_9$; 
the result in Claim~\ref{claim3} narrows the Henneberg-I step that follows
$v_9\triangleleft(v_1, v2)$ to either $v_{10} \triangleleft (v_2, v_9)$
or $v_{10} \triangleleft (v_1, v_9)$, so in Claim~\ref{claim5} we show that
 $v_{10} \triangleleft (v_1, v_9)$ is infeasible; finally
Claim~\ref{claim6} shows that the only remaining possibility namely 
$v_{10} \triangleleft (v_2, v_9)$ results in a consequence  that 
contradicts to Claim~\ref{claim1}. This implies no minimal graph $G$
can exist that contradicts the conditions of the theorem, thus proving
Theorem~\ref{the:quantifierExchange}.
\end{proof}

\section{Conclusions and Future Work}
\label{sec:conclusion}

By studying the configuration spaces of Simple 1-dof Henneberg-I graphs, 
we have taken the next step in a systematic and graded program
laid out in \cite{bib:Gao08} - for the  
combinatorial characterizations of efficient configuration spaces
of underconstrained 2D Euclidean Distance Constraint Systems (resp. frameworks).
In particular, the results presented here go the next step
beyond graphs with connected and convex configuration spaces studied in 
\cite{bib:GaoSitharam08a}.

A generalization of the results presented here
from Henneberg-1 graphs to the larger class of Tree- or 
Triangle-decomposable graphs
appears in \cite{bib:Gao08} and \cite{bib:GaoSitharam08b}.

As immediate future work, it would be desirable to give a cleaner
combinatorial characterization of low sampling complexity for 
configuration spaces of 
1-path Simple 1-dof Henneberg-I graphs. I.e, it would be desirable to 
improve the characterization of Theorem~\ref{the:chain}.
The next natural continuation is to study configuration spaces
of graphs with $k$ dofs ($k>1$) obtained by deleting
$k$ edges from  Henneberg-I  
or Tree- or Triangle-decomposable graphs.


\end{document}